\documentclass[12pt]{article}

\usepackage{amsmath,amsfonts,amssymb,amsbsy}
\usepackage{graphicx,color}
\usepackage{graphics}
\usepackage{cite}

\usepackage{verbatim}
\usepackage{stmaryrd}
\usepackage{amsfonts}

\usepackage{tikz}
\usetikzlibrary{matrix,arrows,shapes}

\newcommand{\be}{\begin{equation}}
\newcommand{\ee}{\end{equation}}
\newcommand{\ba}{\begin{eqnarray}}
\newcommand{\ea}{\end{eqnarray}}
\newcommand{\bea}{\begin{eqnarray}\begin{array}}
\newcommand{\eea}{\end{array}\end{eqnarray}}
\newcommand{\nn}{{\nonumber}}
\newcommand{\bc}{\begin{center}}
\newcommand{\ec}{\end{center}}

\newcommand{\beaa}{\begin{eqnarray}}
\newcommand{\eeaa}{\end{eqnarray}}

\newcommand{\bmx}{\begin{pmatrix}}
\newcommand{\emx}{\end{pmatrix}}

\newcommand{\mf}{\mathfrak}

\newcommand{\alf}{{\textstyle{\frac{1}{2}}}}

\newcommand{\Smat}{{\mathcal S}}
\newcommand{\Refl}{{\mathcal R}}
\newcommand{\Intertwiner}{{\mathcal I}}

\newcommand{\eps}{\epsilon}
\newcommand{\tr}{\,{\rm tr}}

\newcommand{\ket}[1]{{\,\left|#1\right>}\,}

\newcommand{\vac}{\ket 0}

\newcommand{\li}{{\mathrm I}}
\newcommand{\lii}{{\mathrm{II}}}
\newcommand{\liii}{{\mathrm{III}}}
\newcommand{\KI}{{K^\li}}
\newcommand{\KII}{{K^\lii}}
\newcommand{\KIII}{{K^\liii}}
\newcommand{\etap}{\tilde\eta}
\newcommand{\KIIA}{{K_{(\alpha)}^\lii}}
\newcommand{\KIIIA}{{K_{(\alpha)}^\liii}}
\newcommand{\KIIn}[1]{{K_{(#1)}^\lii}}
\newcommand{\KIIIn}[1]{{K_{(#1)}^\liii}}

\newcommand{\LL}{{\mathrm L}}
\newcommand{\RR}{{\mathrm R}}

\def\be{\begin{equation}}
\def\ee{\end{equation}}
\def\ba{\begin{eqnarray}}
\def\ea{\end{eqnarray}}
\def\bc{\begin{center}}
\def\ec{\end{center}}

\def\nn{\nonumber}

\def\r2{{\sqrt{2}}}

\newcommand{\btp}{\begin{tikzpicture}[baseline=-5pt,scale=0.25,line width=0.7pt]}
\newcommand{\etp}{\end{tikzpicture}}

\newcommand{\SIcI}{S^{\li,\,\li}}
\newcommand{\SIIcI}{S^{\lii,\,\li}}
\newcommand{\SIcII}{S^{\li,\,\lii}}

\newcommand{\SIIIcII}{S^{\liii,\,\lii}}
\newcommand{\SIIcIII}{S^{\lii,\,\liii}}
\newcommand{\SIIIcIII}{S^{\liii,\,\liii}}
\newcommand{\RIII}{R^{\liii}}
\newcommand{\RII}{R^{\lii}}
\newcommand{\RI}{R^{\li}}
\newcommand{\fL}{f^L}
\newcommand{\fR}{f^R}
\newcommand{\fB}{f^\tau}
\newcommand{\ff}{f^\sigma}
\newcommand{\ffB}{f^{\sigma\tau}}
\newcommand{\hL}{h^L}
\newcommand{\hR}{h^R}
\newcommand{\xp}{x^{+}}
\newcommand{\xm}{x^{-}}

\newcommand{\xpm}{x^\pm}
\newcommand{\mxmp}{-x^\mp}
\newcommand{\x}{x}
\newcommand{\phasef}{\zeta}
\newcommand{\acoeff}{\mathbb A}
\newcommand{\fcoeff}{\mathbb F}

\begin{document}

\begin{titlepage}
\begin{flushright}
DAMTP-2009-83\\
YITP-09-99\\
\end{flushright}
\begin{centering}
\vspace{.2in}

 {\Large {\bf Asymptotic Bethe equations for open boundaries in planar AdS/CFT}}

\vspace{.3in}

{\large D. H. Correa ${}^{a,1}$ and C. A. S. Young ${}^{b,2}$}\\
\vspace{.2 in}
${}^{a}${\emph{DAMTP, Centre for Mathematical Sciences \\
University of Cambridge\\ Wilberforce Road,
Cambridge CB3 0WA, UK}} \\
\vspace{.2in}
${}^{b}$ {\emph{Yukawa Institute for Theoretical Physics\\
Kyoto University, Kyoto, 606-8502, Japan}}\\

%
\footnotetext[1]{{\tt D.Correa@damtp.cam.ac.uk,}\quad ${}^{2}${\tt cyoung@yukawa.kyoto-u.ac.jp}}
\vspace{.5in}

{\bf Abstract}

\vspace{.1in}

\end{centering}

{
We solve, by means of a nested coordinate Bethe ansatz, the open-boundaries scattering theory describing the excitations of a free open string propagating in $AdS_5\times S^5$, carrying large angular momentum $J=J_{56}$, and ending on a maximal giant graviton whose angular momentum is in the same plane. We thus obtain the all-loop Bethe equations describing the spectrum, for $J$ finite but large, of the energies of such strings, or equivalently, on the gauge side of the AdS/CFT correspondence, the anomalous dimensions of certain operators built using the $\epsilon$ tensor of $SU(N)$.  We also give the Bethe equations for strings ending on a probe D7-brane, corresponding to meson-like operators in an $\mathcal N=2$ gauge theory with fundamental matter.}

\end{titlepage}

\tableofcontents
\section{Introduction}

Following progress in recent years, see e.g. \cite{MZ,BKS,BS,AFS,BKSZ}, the spectral problem in the planar limit of ${\cal N}=4$ super Yang-Mills is nowadays accepted to be integrable. Integrability allows the scale dimensions of very long single-trace operators to be encoded in a certain system of Bethe equations \cite{BSpsu224}. These equations can be derived by solving, by means of a nested Bethe ansatz, the 1+1 dimensional scattering theory  \cite{Bsu22} that describes the excitations about the BPS operator $\tr\left(Z^J\right)$, which serves as the Bethe reference vacuum. Symmetry considerations fix the S matrix of this theory (at least for the elementary particles, cf. \cite{boundstates}) up to an overall scalar factor, which is now also believed to be known \cite{Janik,BHL,BES}. The asymptotic Bethe equations are a key ingredient in the formulation of the TBA equations \cite{TBA} which are believed to encode the spectrum of operators of all lengths.

The spectral problem extends to cases with open boundary conditions. Integrable open boundary conditions appear in the duality between ${\cal N}=4$ $SU(N)$ SYM and IIB strings on $AdS_5\times S^5$ when one considers open strings ending on certain maximal giant gravitons \cite{HMopen}  (for earlier work see \cite{BV,Mann}). These giant gravitons are D3-branes that wrap a maximal $S^3$ of the $S^5$. Such a D3-brane has charge $J=N$ under the angular momentum generator $J$ corresponding to the plane defining the $S^3$, and the dual operator is $\eps^{i_1,\cdots, i_N}_{j_1,\cdots ,j_N} Z^{j_1}_{i_1} \cdots Z^{j_{N}}_{i_{N}}  \sim \det Z$, where $Z$ is the unique scalar field of the ${\cal N}=4$ action with charge $+1$ under $J$. The operator dual to the brane with a single string ending on it has one of these $Z$'s replaced by a chain (i.e. a matrix product) of many ${\cal N} =4$ adjoint fields. To set up an asymptotic scattering theory, one has to pick a Bethe vacuum for this chain -- and in contrast to the closed case, there are inequivalent choices. In \cite{HMopen} it was shown that the scattering theory is integrable for (at least) two choices, whose vacuum states are
\be
\epsilon^{i_1,\cdots, i_N}_{j_1,\cdots ,j_N}
Z^{j_1}_{i_1} \cdots Z^{j_{N-1}}_{i_{N-1}}
(Y^L)^{j_N}_{i_N},
\label{yzeroop}
\ee
and
\be
\epsilon^{i_1,\cdots, i_N}_{j_1,\cdots ,j_N}
Z^{j_1}_{i_1} \cdots Z^{j_{N-1}}_{i_{N-1}}
(\chi_\LL Z^L
 \chi_\RR)^{j_N}_{i_N},
 \label{zzeroopintro}
\ee
with $N\gg L\gg 1$ and where, as we recall below, $\chi_\LL$, $\chi_\RR$ are certain boundary degrees of freedom. The boundary reflection matrices, to all-loop in $\lambda_\text{'t Hooft}$,  were derived for both scattering theories in \cite{HMopen} -- for subsequent progress, see \cite{ChCo,ABR,Ahn:2008df,Murgan:2008zu,Beisert:2008cf,Palla} -- but only for the former has the system been solved, in \cite{Galleas,Nepomechie:2009zi}.

In the present paper our main goal is to fill this gap in the literature by finding the Bethe equations for the latter choice of vacuum, known as the $Z=0$ case. To this end, in section 2 we recall the details of the bulk and boundary scattering theory and proceed to solve it by a nested coordinate Bethe ansatz. In this way we identify the entries of the diagonalised reflection matrix, which allows us to write down the Bethe equations.

In section 3 we go on to find the Bethe equations for a different but closely related set of boundary conditions. These arise when one adds to the gauge theory a chiral hypermultiplet of fundamental matter (breaking ${\cal N}=4$ to ${\cal N}=2$) and uses these fundamental fields to form ``meson-like'' operators $\bar q ZZ\dots ZZq$. This setup is dual to open strings ending on a probe D7-brane \cite{KMMW,ErMa} and the all-loop scattering theory for it was proposed in \cite{CY}.

\section{$Z=0$ giant graviton}
Let $D$ be the dilatation operator, and $J\in so(6)_R$ the R-symmetry which generates rotations in (say) the $56$ plane. Of the (complexified) superconformal algebra $psu(4|4)$ of the $\mathcal N=4$ theory, the subalgebra commuting with $D-J$ is a copy of $psu(2|2)\times psu(2|2)$. The scalar $Z:=\Phi_5 + i\Phi_6$ is the unique field in the $\mathcal N=4$ SYM action whose charge under $D-J$ is zero; on the remaining fields $D-J>0$. There are 16 fields with the smallest positive eigenvalue,  $D-J=1$: the 4 scalars $\Phi_1, \Phi_2, \Phi_3, \Phi_4$, the 4 gauge fields $A_\mu$, and 8 out of the 16 fermions. These fields\footnote{more precisely, cf. eqn (\ref{N4fields}) below.} transform in the bi-fundamental representation $(\boxslash,\boxslash)$ of $psu(2|2)\times psu(2|2)$ and we denote them by $\left\{\chi^{\mathrm a, \dot{\mathrm a}} \right\}_{\mathrm a,\dot{\mathrm a} \in \{1,2,3,4\}}$.

Just as one can build a scattering theory with closed boundaries whose vacuum state is the operator $\tr Z^L$, $L\gg 1$, \cite{Bsu22}, so it is also possible to construct a scattering theory with open  boundary conditions.

In this section the vacuum states will be the operators
\be
\epsilon^{i_1,\cdots, i_N}_{j_1,\cdots ,j_N}
Z^{j_1}_{i_1} \cdots Z^{j_{N-1}}_{i_{N-1}}
(\chi^{\mathrm a,\dot{\mathrm a}}_\LL Z^J
 \chi^{\mathrm e,\dot{\mathrm e}}_\RR)^{j_N}_{i_N},
 \label{zzeroop}
 \ee
where L and R stand for the Left and Right boundaries. The boundary degrees of freedom also transform in a bi-fundamental representation $(\boxslash,\boxslash)$ of $psu(2|2)\times psu(2|2)$ \cite{HMopen} and the full symmetry is preserved by reflections of bulk excitations from the boundary. As usual, we begin by considering the simpler scattering theory in which the symmetry is only one copy of $psu(2|2)$ and both bulk and boundary excitations transform in the fundamental.

\subsection{The scattering theory.}
\paragraph{The $psu(2|2)\ltimes \mathbb R^3$ symmetry algebra.}Recall from \cite{Bsu22,Beis2006} that the algebra $psu(2|2)\ltimes \mathbb R^3$ is spanned by the bosonic generators $\mf R^a{}_b$, $\mf L^\alpha{}_\beta$ of the two $su(2)$ factors and $\mf C,\mf P, \mf K$ of the central extension $\mathbb R^3$, together with the fermionic generators $\mf Q^\alpha{}_a$ and $\mf S^a{}_\alpha$. We write $a,b,\ldots \in \{1,2\}$ and $\alpha,\beta,\ldots \in \{3,4\}$ for fundamental indices of, respectively, the $su(2)_{\mf R}$ and $su(2)_{\mf L}$ factors:
\begin{align} [ {\mf{R}}^{a}_{~{b}}, \mf{J}^{c}]
&= \delta^{c}_{b} \mf{J}^{a}-\tfrac12 \delta^{a}_{b}\mf{J}^{c}\,,
\qquad\; [ {\mf{R}}^{a}_{~{b}}, \mf{J}_{c}]
=-\delta_{c}^{a} \mf{J}_{b}+\tfrac12 \delta^{a}_{b} \mf{J}_{c}\,,\nn\\
[\mf{L}^\alpha_{~\beta},\mf{J}^\gamma]
&=\delta^{\gamma}_\beta \mf{J}^\alpha-\tfrac12\delta^{\alpha}_\beta \mf{J}^\gamma\,,
\qquad [\mf{L}^\alpha_{~\beta},\mf{J}_{\gamma}]
= -\delta_{\gamma}^\alpha \mf{J}_\beta+\tfrac12\delta^{\alpha}_\beta \mf{J}_{\gamma} \,,\end{align}
where $\mf J$ is any generator with the index shown. Then the supersymmetries transform canonically under $su(2)_{\mf L} \times su(2)_{\mf R}$ and close into the bosonic generators according to
\begin{align} \{\mf{Q}^{\alpha}_{~a},\mf{Q}^{\beta}_{~ b}\}
&= \epsilon^{\alpha\beta}\epsilon_{ab}\mf{P} \,,\qquad \quad \
\{\mf{S}^{a}_{~\alpha},\mf{S}^{b}_{~ \beta}\} = \epsilon_{\alpha\beta}\epsilon^{ab}\mf{K} \,,\nn\\
\{\mf{S}^{a}_{~\alpha},\mf{Q}^{\beta}_{~ b}\}
&= \delta^{ a}_{ b} \mf{L}^\beta_{~\alpha} +  \delta^\beta_{\alpha}  {\mf{R}}^{ a}_{~ b}
+ \delta^{ a}_{ b} \delta^\beta_{\alpha} \mf{C} \,.\end{align}

\paragraph{Fundamental representations.}The boundary degrees of freedom and the elementary excitations propagating in the bulk all transform in fundamental representations of $psu(2|2)\ltimes \mathbb R^3$. The carrier space  $(\bf 2|2)$ of such representations has a basis $\ket{\chi^{\mathrm a}} = \{\ket{\phi^a} , \ket{\psi^\alpha}\}$ consisting of an $su(2)_{\mf R}$ doublet of bosons $\ket{\phi^a}$ and an $su(2)_{\mf L}$ doublet of fermions $\ket{\psi^\alpha}$. A fundamental representation is specified by the values of the coefficients $a,b,c,d$ that determine the action of the supersymmetries on these basis states:
\beaa & \mf Q^\alpha{}_a \ket{\phi^b} = a \delta_a^b \ket{\psi^\alpha}, \qquad\;\; &
    \mf Q^\alpha{}_a \ket{\psi^\beta} = b \eps^{\alpha\beta} \eps_{ab} \ket{\phi^b}, \\
&\mf S^a{}_\alpha \ket{\phi^b} = c \eps_{\alpha\beta} \eps^{ab} \ket{\psi^\beta}, \quad &
    \mf S^a{}_\alpha \ket{\psi^\beta} = d \delta_{\alpha}^\beta \ket{\phi^a}, \eeaa
which must obey the shortening (or mass-shell) condition $ad-bc=1$.

For an elementary magnon propagating in the bulk, with momentum $p$, these coefficients are parameterized as~\cite{Beis2006}
\be a=\sqrt{g}\, \eta,\quad b= -\sqrt{g}\,\frac{i\phasef}{\eta}\left(1-\frac{x^+}{x^-}\right),\quad
c=- \sqrt{g}\,\frac{\eta}{\phasef x^+},\quad d=\sqrt{g}\,\frac{x^+}{i \eta}\left(1-\frac{x^-}{x^+}\right), \label{qn} \ee
where $|\eta|^2=i(x^--x^+)$, to ensure unitarity, $x^\pm$ are the standard spectral parameters obeying
\be e^{i p} =  \frac{x^+}{x^-}, \qquad
x^+ +\frac{1}{x^+}-x^- -\frac{1}{x^-}=\frac{i}{g}, \label{sp}\ee
and $\phasef$ is pure phase given by the product $\prod_k e^{ip_k}$ over all magnons to the left of the magnon in question. The values of the central charges $\mf{C,P,K}$ are given in terms of $p$ and $\phasef$ by
\beaa
&& P = ab = g\phasef\left(1- e^{ip} \right) ,\qquad K = cd = \frac{g}{\phasef}\left(1-e^{-ip}\right), \label{PK}
\\
&& C = \alf(ad+bc)= \alf\sqrt{1+16g^2 \sin(\tfrac{p}{2})^2},\label{C}\eeaa
and the energy $E$ of the magnon is identified with $2C$. We will write $\boxslash_{(p,\phasef,\eta)}$ to denote this representation.

The boundary excitations, on the other hand, do not carry momentum. They transform in the representation given by  \cite{HMopen}
\be a_B = \sqrt g \,\eta_B, \quad                   b_B =-\sqrt g\,\frac{ i \phasef_B}{\eta_B} ,\quad
    c_B = -\sqrt g \,\frac{ \eta_B}{x_B \phasef_B}, \quad d_B= \sqrt g\,\frac{ x_B}{i \eta_B}, \label{bndryabcd}\ee
where $\left|\eta_B\right|^2=-ix_B$, $\phasef_B$ is a boundary phase to be specified below, and the mass-shell condition $ad-bc=1$ now reads
\be x_B + \frac{1}{x_B} = \frac{i}{g} \,.\label{xBeqn}\ee
The values of the central charges $\mf{C,P,K}$ and the energy $E$ of an unexcited boundary are given by
\be P = a_B b_B = g i\phasef_B, \qquad K= \frac{g}{i \phasef_B}, \qquad \alf E = C = \alf\sqrt{1+4g^2} . \ee
We write this representation as $\boxslash_{(\phasef_B,\eta_B)}$.

\paragraph{Bulk and boundary scattering.}Asymptotic components of energy eigenstates transform in tensor products of these representations,
\be \boxslash_{(\phasef_L,\eta_L)} \otimes \boxslash_{(p_1,\phasef_1,\eta_1)} \otimes \dots \otimes \boxslash_{(p_\KI,\phasef_\KI,\eta_\KI)} \otimes \boxslash_{(\phasef_R,\eta_R)} \label{ascmpt}\ee
where $\KI$ is the number of bulk magnons.
The phases $\phasef_L,\phasef_R$ and $\phasef_i$ associated to all the particles, bulk and boundary, are conveniently visualized using the Lin-Lunin-Maldacena (LLM) disk picture \cite{HM,LLM}. In this picture, the boundary degrees of freedom correspond to radial line segments, and bulk excitations to line segments between points on the circumference. For example, an asymptotic component of a state with three bulk magnons might look as follows \cite{HMopen}.
\be \btp\begin{scope}[>=latex,scale=.8]
     \draw[->] (0,0) --  (20:4.5); \draw (20:4) --  (20:8) node [above right] {$\phasef_1 = - \phasef_L$};
     \draw[->] (20:8) -- (-3:7.527);
     \draw (20:8) -- (-20:8) node [right] {$\phasef_2 = \phasef_1 e^{ip_1}$};
     \draw[->](-20:8) -- (-87:3.39);
     \draw (-20:8) -- (-150:8) node [below left] {$\phasef_3 = \phasef_2 e^{ip_2}$} ;
     \draw[->] (-150:8) -- (-195:5.907);
     \draw (-150:8) -- (-235:8)  node [above left] {$\phasef_R = \phasef_3 e^{ip_3}$};
     \draw[->] (-235:8) -- (-235:3.5);
     \draw (-235:4) -- (0,0);
     \draw (0,0) circle (8);
     \filldraw[black] (0,0) circle (.3);
  \end{scope}
\etp\nn
\ee

As usual in 1+1 dimensional scattering theories with boundaries, an asymptotic region is labelled by the ordering of the bulk particles (specified by a permutation $\sigma\in S_\KI$ of some fiducial ordering) \emph{and} a sign $\pm 1$ for each bulk particle which specifies whether it is ingoing or outgoing from (say) the right boundary. That is, the asymptotic regions correspond to the Weyl chambers of the $BC_\KI\equiv S_\KI \ltimes \mathbb Z_2^\KI$ group of reflections \cite{gaudin,olpe,che,Sutherland}. The components of an energy eigenstate in different asymptotic regions are related by the bulk and boundary scattering matrices, $\Smat$ and $\Refl$; to respect the symmetry of the problem, $\Smat$ and $\Refl$ must commute with the action of $psu(2|2)\ltimes \mathbb R^3$. The labels of the representations can change under scattering, but must do so in a way which preserves the values of the three central charges $\mf{C,P,K}$. The correct changes turn out to be \cite{Bsu22,Beis2006,HMopen}
\beaa
\label{sch}
\Smat:\boxslash_{(p,\phasef,\eta)} \otimes \boxslash_{(p',\phasef e^{ip},\eta')}
&\longrightarrow&   \boxslash_{(p',\phasef,\etap')}\otimes \boxslash_{(p,\phasef e^{ip'},\etap)} \\
    \Refl_L:   \boxslash_{(-\phasef,\eta_B)}         \otimes \boxslash_{(p,\phasef,\eta)} &\longrightarrow&
           \boxslash_{(-\phasef e^{2ip},\etap_B)}\otimes \boxslash_{(-p,\phasef e^{2ip},\etap)} \\
    \Refl_R:   \boxslash_{(p,\phasef,\eta)}          \otimes \boxslash_{(\phasef e^{ip},\eta_B)}&\longrightarrow&
               \boxslash_{(-p,\phasef,\etap)}   \otimes \boxslash_{(\phasef e^{-ip},\etap_B)} \eeaa
which are rather natural when visualized in the LLM disk picture:
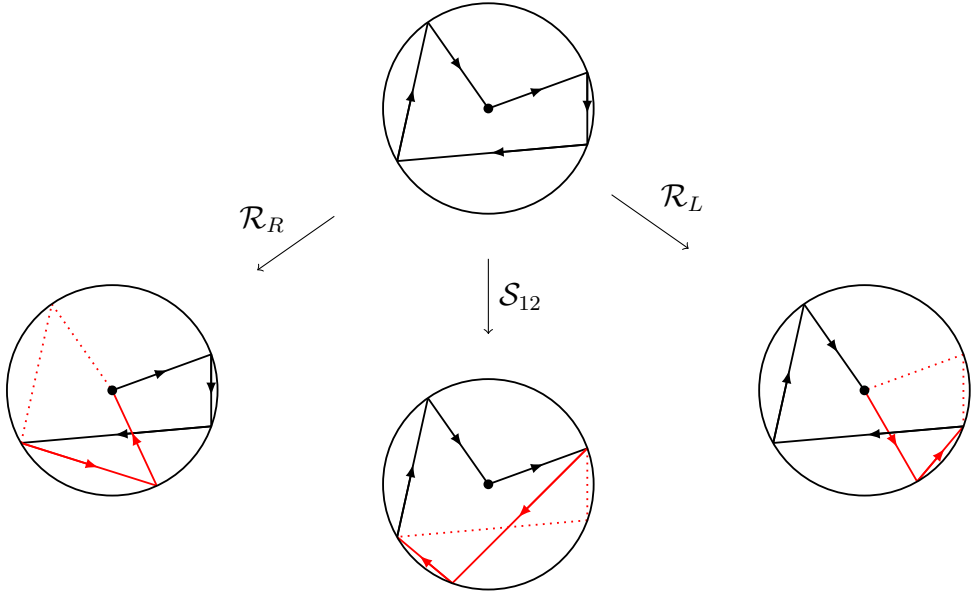
\be \btp
\begin{scope}[>=latex,scale=.7]
\draw[->] (0,0) --  (20:4.5);  \draw (20:4) --  (20:8);
\draw[->] (20:8) -- (-3:7.527); \draw (20:8) -- (-20:8);
\draw[->](-20:8) -- (-87:3.39); \draw (-20:8) -- (-150:8);
\draw[->] (-150:8) -- (-195:5.907); \draw (-150:8) -- (-235:8);
\draw[->] (-235:8) -- (-235:3.5); \draw (-235:4) -- (0,0);
     \draw (0,0) circle (8);
     \filldraw[black] (0,0) circle (.3);
\end{scope}
\begin{scope}[>=latex,xshift=-20cm,yshift=-15cm,scale=.7]
\draw[->] (0,0) --  (20:4.5);  \draw (20:4) --  (20:8);
\draw[->] (20:8) -- (-3:7.527); \draw (20:8) -- (-20:8);
\draw[->](-20:8) -- (-87:3.39); \draw (-20:8) -- (-150:8);
\draw[color=red] (-150:8) -- (-65:8); \draw[color=red,->] (-150:8) -- (-100:5.95);
\draw[->,color=red] (-65:8) -- (-65:3.5); \draw[color=red] (-65:4) -- (0,0);
\draw[dotted,color=red] (-235:8) --  (0,0);
\draw[dotted,color=red] (-150:8) -- (-235:8);
\draw (0,0) circle (8);
     \filldraw[black] (0,0) circle (.3);
\end{scope}
\begin{scope}[>=latex,xshift=0cm,yshift=-20cm,scale=.7]
\draw[->] (0,0) --  (20:4.5);  \draw (20:4) --  (20:8);
\draw[color=red] (20:8) -- (-110:8); \draw[->,color=red] (20:8) -- (-48:3.38);
\draw[color=red] (-110:8) -- (-150:8);\draw[->,color=red] (-110:8) -- (-134:7.55);
\draw[dotted,color=red] (-20:8) -- (-150:8);
\draw[dotted,color=red] (20:8) -- (-20:8);
\draw[->] (-150:8) -- (-195:5.907); \draw (-150:8) -- (-235:8);
\draw[->] (-235:8) -- (-235:3.5); \draw (-235:4) -- (0,0);
    \draw (0,0) circle (8);
     \filldraw[black] (0,0) circle (.3);
\end{scope}
\begin{scope}[>=latex,xshift=20cm,yshift=-15cm,scale=.7]
\draw[->,color=red] (0,0) -- (-60:5); \draw[color=red] (-60:4) -- (-60:8);
\draw[color=red](-60:8) -- (-20:8); \draw[->,color=red] (-60:8) -- (-36:7.55);
\draw[->](-20:8) -- (-87:3.39); \draw (-20:8) -- (-150:8);
\draw[->] (-150:8) -- (-195:5.907); \draw (-150:8) -- (-235:8);
\draw[->] (-235:8) -- (-235:3.5); \draw (-235:4) -- (0,0);
\draw[dotted,color=red] (20:8) -- (-20:8);
\draw[dotted,color=red] (20:8) -- (0,0);
     \draw (0,0) circle (8);
     \filldraw[black] (0,0) circle (.3);
\end{scope}
\draw[thin,->] (-145:10) -- node[above left] {$\Refl_R$}  ++(-145:5);
\draw[thin,->] (-35:8) -- node[above right] {$\Refl_L$}  ++(-35:5);
\draw[thin,->] (-90:8) -- node[right] {$\Smat_{12}$} ++(-90:4);
\etp\nn
\ee
The tensor product of two fundamental representations is irreducible for generic values of the parameters, and therefore (by Schur's lemma) each of the maps $\Smat_{12}$, $\Refl_L$ and $\Refl_R$ is determined by symmetry up to an overall factor. The most general intertwiner $\Intertwiner$ of $su(2)\oplus su(2)$ representations is
\beaa \Intertwiner \ket{\phi^a \phi^b} &=& A \ket{\phi^{\{a} \phi^{b\}}}
                                       +B \ket{\phi^{ [a} \phi^{b ]}}
     + \alf C \eps^{ab} \eps_{\alpha\beta}\ket{\psi^\alpha\psi^\beta} \nn\\
      \Intertwiner \ket{\psi^\alpha \psi^\beta} &=& D \ket{\psi^{\{\alpha} \psi^{\beta\}}}
                                       +E \ket{\psi^{ [\alpha} \psi^{\beta ]}}
     + \alf F \eps_{ab} \eps^{\alpha\beta}\ket{\phi^a\phi^b} \nn\\
      \Intertwiner \ket{\phi^a \psi^\beta} &=& G \ket{\psi^\beta \phi^a}
                                       +  H \ket{\phi^a \psi^\beta } \nn\\
      \Intertwiner \ket{\psi^\alpha \phi^b} &=& K \ket{\psi^\alpha \phi^b}
                                       +  L \ket{\phi^b \psi^\alpha }\,,\nn\eeaa
for some coefficients $A,B,C,D,E,F,G,H,K,L$, which are then fixed by demanding that $\Intertwiner$ commute with the supersymmetries. They were computed in \cite{Bsu22} for the bulk S matrix and \cite{HMopen} for the boundary reflection matrix, and are reproduced in tables \ref{fig:bulkS} and \ref{fig:bndryRright}. Note that we have not yet specified the parameters $\eta$, and for the moment we allow them to change in an unspecified way $\eta\to\etap$ under scattering.
\begin{table}
{\small
\beaa A &=& S_0(p_1,p_2) \frac{\eta_1\eta_2}{\etap_1\etap_2}
                                  \frac{x_2^+- x_1^-}{x_2^- - x_1^+} \nn\\
      B &=& S_0(p_1,p_2) \frac{\eta_1\eta_2}{\etap_1\etap_2}
                              \frac{x_2^+ - x_1^-}{x_2^- - x_1^+}
            \left( 1 - 2 \frac{1 - 1/x_2^- x_1^+}{1 - 1/x_2^+ x_1^+}\,\,
                         \frac{ x_2^- - x_1^-}{x_2^+ - x_1^-}\right) \nn\\
C &=&  -\frac{2i\eta_1\eta_2}{\phasef} S_0(p_1,p_2) \frac{1}{ x_1^+ x_2^+} \frac{ x_2^- - x_1^-}{x_2^- - x_1^+} \frac{1}{1 - 1/x_2^+ x_1^+} \nn\\
D &=& - S_0(p_1,p_2)\nn\\
E &=& - S_0(p_1,p_2)\left( 1 -  2 \frac{ 1 - 1/x_2^+ x_1^-}{1 - 1/x_2^- x_1^-}\,\,
                             \frac{ x_2^+ - x_1^+}{x_2^- - x_1^+}\right)\nn\\
F &=& -\frac{2i\phasef}{\etap_1\etap_2}
       S_0(p_1,p_2) \frac{(x_1^+ - x_1^-) (x_2^+ - x_2^-)}{ x_1^- x_2^-}\,
         \frac{x_2^+ - x_1^+}{x_2^- - x_1^+}\,\frac{1}{1 - 1/x_2^- x_1^-} \nn\\
G &=&  S_0(p_1,p_2)\frac{\eta_1}{\etap_1}\frac{x_2^+ - x_1^+}{x_2^- - x_1^+}
\qquad
H \ = \ S_0(p_1,p_2)\frac{\eta_1}{\etap_2} \frac{ x_2^+ - x_2^-}{ x_2^- - x_1^+} \nn
\\
K &=&  S_0(p_1,p_2)\frac{\eta_2}{\etap_1} \frac{ x_1^+ - x_1^-}{ x_2^- - x_1^+}
\qquad
L \ = \ S_0(p_1,p_2)\frac{\eta_1}{\etap_1}\frac{x_2^- - x_1^-}{x_2^- - x_1^+}\,. \nn
\eeaa}
\vspace{-1cm}
\caption{{\small Coefficient functions for the bulk scattering matrix of two elementary magnons.}}
\label{fig:bulkS}
\end{table}
\begin{table}
{\small
\beaa A &=&  R_0(p) \frac{\eta_B\eta}{\etap_B\etap}\frac{x^- \left(x^--x_B\right)}{x^+ \left(x^++x_B\right)}\,,\nn\\
      B &=&   R_0(p) \frac{\eta_B\eta}{\etap_B\etap}\frac{x^- \left(-2 \left(x^-\right)^2+x^+
   x^-+2 \left(x^+\right)^2\right)-x_B \left(2 \left(x^-\right)^2+x^+ x^--2
   \left(x^+\right)^2\right)}{\left(x^+\right)^2 \left(x_B+x^+\right)}\,,\nn\\
      C &=& - R_0(p) \frac{2i\eta_B\eta}{\phasef} \frac{
   \left(x_B+x^--x^+\right) \left(x^-+x^+\right) }{ x^+
   \left(x_B+x^+\right)}\,,\nn\\
      D &=& R_0(p)\,, \nn\\
      E &=& R_0(p)\frac{x^+ \left(2 \left(x^-\right)^2+x^+ x^--2 \left(x^+\right)^2\right)+x_B
   \left(-2 \left(x^-\right)^2+x^+ x^-+2 \left(x^+\right)^2\right)}{x^- x^+ \left(x_B+x^+\right)}\,,\nn\\
      F &=& -R_0(p) \frac{2i\phasef}{\etap_B \etap} \frac{
\left(\left(x^-\right)^2-\left(x^+\right)^2\right) \left(x^- x^++x_B \left(x^+-x^-\right)\right)}{x^- \left(x^+\right)^2 \left(x_B+x^+\right)}\,,\nn\\
      G &=&  - R_0(p)\frac{\eta}{\etap_B} \frac{x_B
   \left(x^-+x^+\right)}{ x^+\left(x_B+x^+\right)}\,,
   \qquad
    H \ = \ R_0(p)\frac{\eta}{\etap}\frac{\left(x^+\right)^2-x_B x^-}{x^+
   \left(x_B+x^+\right)}\,,\nn
   \\
    K &=& R_0(p)\frac{\eta_B}{\etap_B}\frac{\left(x^-\right)^2+x_B x^+}{\left(x^+\right)^2+x_B
   x^+}\,,
     \qquad
     L \ = \  R_0(p) \frac{\eta_B}{\etap} \frac{\left(x^--x^+\right) \left(x^-+x^+\right)}{x^+ \left(x_B+x^+\right) }\,,\nn
\eeaa}
\vspace{-1cm}
\caption{{\small Coefficient functions for the right reflection of an elementary magnon. Left reflection ones are obtained
by parity symmetry. In the LLM disk, this is visualized by reversing the arrows, i.e. $x^{\pm} \to -\x^{\mp}$ and $\phasef \to -\phasef \tfrac{x^+}{x^-}$.}}
\label{fig:bndryRright}
\end{table}

\subsection{Coordinate Bethe Ansatz}
\label{CBAgg}
We can now turn to solving the scattering problem by Bethe ansatz methods. As usual when treating integrable systems with boundaries, the strategy is to begin by considering the scattering problem on the half-line with one boundary.  One uses a Bethe ansatz to construct (the asymptotic components of) energy eigenstates for this semi-infinite system, parameterized by a collection of continuous parameters (the particle rapidities). Then the next step is to introduce the other boundary, which will place extra consistency conditions (the Bethe equations) on the rapidities -- thereby quantizing the spectrum, as one expects for a system in finite volume.

Following the work of Sklyanin \cite{Sklyanin}, systems with boundaries are very commonly treated by means of the algebraic Bethe ansatz \cite{FaddeevABA}. This was the approach taken in \cite{Galleas} for $Y=0$ giant graviton boundary conditions. But it is certainly also possible to use a coordinate Bethe ansatz in systems with boundary: see \cite{gaudin,Sutherland}, and, for a system (the Hubbard model) which requires nesting  \cite{schulz}. We adopt the coordinate approach here because, although it is perhaps less mathematically deep, its physical interpretation is slightly more transparent; and our goal is to obtain the Bethe equations with the minimum of effort. We shall follow rather closely the notation used in \cite{Bsu22} in solving the closed case.

Let us, then, consider the scattering problem on the half-line with, say, a right boundary.  Consider states with $\KI$ elementary bulk particles. An asymptotic component
\be\ket{\chi_1^{\mathrm a}}\otimes
          \dots \otimes\ket{\chi_\KI^{\mathrm y}}\otimes \ket{\chi_R^{\mathrm z}} \in
\boxslash_{(p_1,\phasef_1,\eta_1)} \otimes \dots \otimes \boxslash_{(p_\KI,\phasef_\KI,\eta_\KI)} \otimes \boxslash_{(\phasef_R,\eta_R)} \ee
of such a state can be abbreviated as
\be \ket{\chi_1^{\mathrm a}\,
                        \dots \,\chi_\KI^{\mathrm y}\, \chi_R^{\mathrm z}}^\li.\label{bsac}\ee
Any such asymptotic component extends, in a unique way, to an energy eigenstate:  the components in the remaining asymptotic regimes are obtained by acting with all possible products of
\be \Smat^\li_{12}, \Smat^\li_{23},\dots, \Smat^\li_{\KI-1,\,\KI}\quad\text{ and } \quad \Refl^\li.\ee
(Here we have introduced the superscript $^\li$ to distinguish these as the \emph{level $\li$} states and scattering operators.)
In general, however, the internal indices $\mathrm a,\mathrm b,\dots$ of the particles will change in a complicated way under these scattering operations. The \emph{nested coordinate Bethe ansatz} \cite{yang} consists in choosing a special subspace of states $\ket{\Psi}$ on which, by contrast, $\Smat^\li_{i,\,i+1}$ and $\Refl^\li$ act merely by changing the representation labels (as discussed above) and multiplying by fixed \emph{scalar} factors $\SIcI_{i,\,i+1}$ and $\RI$. On such states $\ket\Psi$ the theory is, loosely speaking, as close as possible to one with diagonal scattering. Precisely, we demand
\be \Smat_{i,\,i+1} \ket{\Psi} = \ket{\Psi}_{\sigma_{i,\,i+1}} \SIcI_{i,\,i+1},\qquad
    \Refl \ket{\Psi} = \ket{\Psi}_{\tau} \RI\label{BetheStateI}\ee
where
\be \sigma_{12}, \sigma_{23}, \dots, \sigma_{\KI-1,\,\KI},  \tau \ee
are the operators which change the representation labels (thus mapping to a ket in a neighbouring asymptotic region) but which leave unaltered the internal indices $\mathrm a, \mathrm b, \dots$ of the basis states (\ref{bsac}). They obey the defining relations
\be\begin{split} \sigma_{i,\,i+1} \sigma_{i+1,\,i+2} \sigma_{i,\,i+1} =  \sigma_{i+1,\,i+2}  \sigma_{i,\,i+1} \sigma_{i+1,\,i+2},\qquad \sigma_{i,\,i+1}^2=\mathrm{id} \\ \tau^2=\mathrm{id},\qquad
\tau\, \sigma_{\KI-1,\,\KI}\,\tau\, \sigma_{\KI-1,\,\KI} =  \sigma_{\KI-1,\,\KI}\,\tau\, \sigma_{\KI-1,\,\KI}\,\tau.\end{split}\label{BCK}\ee
of the $BC_{\KI}$ group. The $\Smat_{i,\,i+1}$ and $\Refl$ also realize these relations, which is really the precise statement of integrability here: it is what guarantees that the extension from one asymptotic region to all the others can be consistently completed by adding only a finite number of terms to the state vector.  Note that, here, $\Smat_{i,\,i+1}$ is the scattering of the $i$th and $(i+1)$st particles \emph{labelled as they are ordered in space}, and consequently it is the braided version of the Yang-Baxter equation, which is the first of the relations in (\ref{BCK}), that the $\Smat_{i,\,i+1}$ obey.

\subsubsection*{Level II}
We first define the \emph{level $\lii$} vacuum to be the state
\be \vac^\lii := \ket{\psi_1^3\psi_2^3 \dots \psi_\KI^3 \psi_R^3}.\ee
This is an $su(2)\oplus su(2)$ highest-weight state, so indeed $\Smat^\li_{i,i+1}$ and $\Refl^\li$ can only act diagonally. From tables \ref{fig:bulkS} and \ref{fig:bndryRright} one sees that
\be  \SIcI_{i,\,i+1} = -1, \qquad  \RI  =1.\ee

\paragraph{Single particles: bulk.}The next step is to define additional states -- interpreted as \emph{level $\lii$ excitations} above this level~$\lii$ vacuum -- with the property that they transform under $\Smat^\li_{i,i+1}$ and $\Refl^\li$ in exactly the same fashion as $\vac^\lii$. Consider first single excitations, and temporarily forget about the boundary. The situation is then just as in \cite{Bsu22}: one makes a spin-wave ansatz
\be \ket{\phi^a(y)}^\lii_\text{left tail} := \sum_{k=1}^\KI \ket{\psi_1^3\psi_2^3 \dots \phi_k^a \dots\psi_\KI^3}
                 \prod_{\ell=1}^{k-1} \SIIcI(y;\x_\ell) \fL(y;\x_k,\eta_k) .\ee
Here it is necessary to include the ``tail'' running to the left of the particle because the background is inhomogeneous. The level $\li$ parameters (i.e. the representation labels $\xpm_i$, $\phasef_i$ and $\eta_i$) have the status of inhomogeneities at the sites of the level $\lii$ spin chain, and \emph{a priori} $\fL$ and $\SIIcI$ can depend on all of them, though in fact they need only depend on the arguments shown. It suffices to consider a chain of length $\KI=2$. The compatibility condition is then
\be \Smat_{12}^\li \ket{\phi^a(y)}^\lii = \ket{\phi^a(y)}^\lii_{\sigma_{12}}\, \SIcI =-\ket{\phi^a(y)}^\lii_{\sigma_{12}} .\label{cc1}\ee
One finds a solution\footnote{Readers familiar with the literature \cite{AFZ,AF2,deLeeuw,Martins} will note that here $\SIIcI$ does not include $\sqrt{\xp/\xm}$ factors, and may object that we should be using the ``string basis'' for the $\eta$ and $\etap$ parameters in order to produce them. So we should stress that this solution, though not unique, is valid for any choice of $\eta$'s (constrained only by the requirement that $\Smat_{i,\,i+1}$ and $\Refl$ realize (\ref{BCK})).
This is so because, possibly unusually, we chose to treat the $\eta_i$ as level $\li$ parameters on the same footing as the $\xpm_i$. Both $\fL$ and $\SIIcI$ can thus depend explicitly on $\eta$, just as on $\xpm$, and this is reflected in the form of the compatibility condition (\ref{cc1}).
With the shorthand $F_i =\eta_i \fL(y;\x_i,\eta_i)$, $\widetilde F_i =\etap_i \fL(y;\x_i,\etap_i)$, $S_i=  \SIIcI(y;\x_i,\eta_i)$ and  $\widetilde S_i=  \SIIcI(y;\x_i,\etap_i)$, one finds that (\ref{cc1}) unpacks to give
\beaa (\xm_2-\xp_2) F_1 +
      (\xm_1-\xm_2) F_2 S_1
  &=&-(\xp_1-\xm_2) \widetilde F_2 \\
      (\xp_1-\xp_2) F_1 +
      (\xm_1-\xp_1) F_2 S_1
  &=&-(\xp_1-\xm_2) \widetilde F_1 \widetilde S_2 \eeaa
and hence
\be F_1 +  \widetilde F_1 \widetilde S_2 = \widetilde F_2 +F_2 S_1.\ee
The equation above is separable if $\widetilde F_1 = a_2 F_1$ and $F_2 = a_1 \widetilde F_2$ for some function $a_i=a(\x_i,f_i,\eta_i)$. We are quite free to take the simplest possibility, namely $a\equiv 1$, yielding the solution shown in the text. Thus, for us, $\sqrt{\xp/\xm}$ factors do not originate in the choice of $\eta$'s, and we shall introduce them by different reasoning in \S \ref{Beqns} below. }
\be \fL(y;\x,\eta) = \frac{1}{\eta} \frac{\xp-\xm}{y-\xm}, \qquad
    \SIIcI(y;\x) = - \frac{y-\xp}{y-\xm}.\ee
It is useful to define, in addition,
\be \SIcII(\x;y) =  1\big/ \SIIcI(y;\x)=  - \frac{y-\xm}{y-\xp}\label{sicii}\ee
\be \fR(\x,\eta;y) =  \SIcII(\x;y) \fL(y;\x,\eta) = \frac{1}{\eta} \frac{\xm-\xp}{y-\xp}\label{fr}\ee
and verify that the compatibility condition (\ref{cc1}) is also solved by the spin-wave with its tail trailing away to the right,
\be \ket{\phi^a(y)}^\lii_\text{right tail} := \sum_{k=1}^\KI \ket{\psi_1^3\psi_2^3 \dots \phi_k^a \dots\psi_\KI^3}
                 \prod_{\ell=k+1}^{\KI} \SIcII(\x_\ell;y) \fR(\x_k,\eta_k;y) .\label{righttail}\ee

\paragraph{Single particles: boundary} Let us now re-introduce the boundary. One certainly expects that a Bethe state with a single level $\lii$ excitation should be a linear combination of an ingoing (right-moving) spin-wave, an outgoing (left-moving) spin-wave, and a term in which the excitation has just reached the boundary. The subtlety is in arranging the tails consistently, but the correct answer is easy to guess pictorially:
\beaa \ket{\Psi^\lii_{(y,a)}} &=& \dots +
   \btp \draw[very thick] (3,-2) -- (3,2);
        \draw[dotted] (2,-2) -- (2,2);
        \draw[dotted] (1,-2) -- (1,2);
        \draw[->] (0,-2)   -- (2,1);\etp \,\,\,
+  \btp \draw[very thick] (3,-2) -- (3,2);
   \draw[dotted] (2,-2) -- (2,2);
 \draw[dotted] (1,-2) -- (1,2);
        \draw[->] (0,-2)  -- (3,1);\etp\,\,\,
+  \btp \draw[very thick] (3,-2) -- (3,2);
   \draw[dotted] (2,-2) -- (2,2);
 \draw[dotted] (1,-2) -- (1,2);
        \draw[->] (0,-2)  -- (3,0) -- (2,1);\etp\,\,\,
 + \cdots \label{fullIIone}\\
 &=&  \sum_{k=1}^\KI \ket{\psi_1^3\psi_2^3 \dots \phi_k^a \dots\psi_\KI^3\psi_R^3}
                 \prod_{\ell=1}^{k-1} \SIIcI(y;\x_\ell) \fL(y;\x_k,\eta_k) \nn\\
 &&{}+\ket{\psi_1^3\psi_2^3 \dots \dots\psi_\KI^3\phi_R^a}
                 \prod_{\ell=1}^{\KI} \SIIcI(y;\x_\ell) \fB(y;x_B,\eta_B) \nn\\
 &&{}+ \sum_{k=1}^\KI \ket{\psi_1^3\psi_2^3 \dots \phi_k^a \dots\psi_\KI^3\psi_R^3}
                 \prod_{\ell=1}^{\KI} \SIIcI(y;\x_\ell) \RII(y;x_B) \nn\\
 &&\qquad\qquad\qquad\qquad\qquad\qquad
     \times\prod_{\ell=k+1}^{\KI} \SIcII(\x_\ell;-y) \fR(\x_k,\eta_k;-y) \nn\eeaa
for new unknown functions $\fB$ and $\RII$. By construction, this automatically satisfies the compatibility condition everywhere in the bulk. The new compatibility condition is
\be \Refl^\li {\ket{\Psi^\lii_{(y,a)}}} = {\ket{\Psi^\lii_{(y,a)}}}_{\tau}\, \RI
                                   = {\ket{\Psi^\lii_{(y,a)}}}_{\tau} .\label{cc2}\ee
To solve it, it suffices to consider a level $\li$ state with only $\KI=1$ bulk excitation, in which case
\be
{\ket{\Psi^\lii_{(y,a)}} = \mathbb{B} \ket{\psi^3\phi_R^a} + \mathbb{D} \ket{\phi^a\psi_R^3}},
\ee
where we have introduced the shorthands
{\beaa
\mathbb{B}&=&
\btp
\draw[very thick] (3,-2) -- (3,2);
\draw[dotted] (2,-2) -- (2,2);
\draw[->] (1,-2)   -- (2,1);
\etp \,\,\,
+
\btp
\draw[very thick] (3,-2) -- (3,2);
\draw[dotted] (2,-2) -- (2,2);
\draw[->] (1,-2)  -- (3,0) -- (2,1);\etp \,\,\,
=  \fL(y;x,\eta) + \SIIcI(y;x) \RII(y;x_B) \fR(x,\eta;-y),
\\
\mathbb{D} & =&
\btp
\draw[very thick] (3,-2) -- (3,2);
\draw[dotted] (2,-2) -- (2,2);
\draw[->] (1,-2)  -- (3,1);
\etp
\,\,\,
=\SIIcI(y; x) \fB(y;x_B,\eta_B).
\eeaa}
The compatibility condition is then
\ba
K_R(x) \mathbb{D} + G_R(x) \mathbb{B} &=& D_R(x) (\mathbb{D})_\tau ,
\\
L_R(x) \mathbb{D} + H_R(x) \mathbb{B} &=& D_R(x) (\mathbb{B})_\tau ,
\ea
which admits the solution
\be \fB{(y;x_B)} = \frac{1}{\eta_B} \frac{2x_B}{y+x_B},\qquad
    \RII(y;x_B)       = - \frac{y-x_B}{y+x_B},\label{fBRIIsoln}.\ee
The fact that these indeed depend solely on the level $\lii$ rapidity $y$ and the \emph{boundary} level~$\li$ parameters confirms that the ansatz was suitable.

\paragraph{Two particles: bulk.}We now turn to states with $\KII>1$ level $\lii$ excitations.
\emph{}Let us once more temporarily ignore the boundary. An asymptotic component of a level $\lii$ state of two particles, with both tails running to the left, is
\beaa \ket{\phi^a(y_1) \phi^b(y_2)}^\lii_\text{tails left} &=& \sum_{\underset{k<m}{k,m=1}}^\KI \ket{\psi_1^3 \dots \phi_k^a \dots\phi_m^b\dots \psi_\KI^3}
      \prod_{\ell=1}^{k-1} \SIIcI(y_1;\x_\ell) \fL(y_1; \x_k, \eta_k) \nn
\\ &&\qquad\qquad\qquad\qquad\qquad\times
      \prod_{n=1}^{m-1}    \SIIcI(y_2;\x_n)    \fL(y_2;\x_n,\eta_n) \label{bothleft}. \eeaa
The complete level $\lii$ eigenstate of $\KII=2$ particles in the absence of boundaries is (as in \cite{Bsu22}, except that we are working in the ``non-local'' picture without markers $\mathcal Z^\pm$, c.f. \cite{Beis2006})
\beaa \ket{\Psi^\lii_{(y_1,a;\, y_2,b)}} &=&  \ket{\phi^a(y_1) \phi^b(y_2)}^\lii \label{IIbulk}\\
 &&{} + M(y_1,y_2) \ket{\phi^a(y_2) \phi^b(y_1)}^\lii
      + N(y_1,y_2) \ket{\phi^b(y_2) \phi^a(y_1)}^\lii  \nn\\
 &&{} + \eps^{ab} \ket{\psi^4(y_1,y_2)}^\lii.\nn \eeaa
Here the second line is the most general $su(2)$-covariant level $\lii$ scattering matrix.\footnote{Just as at level $\li$, in the presence of a boundary the regions at level $\lii$ correspond to the Weyl chambers of the reflection group $BC_\KII$, and components in different regions are related by scattering operators $\Smat_{i,\,i+1}^\lii$ and $\Refl^\lii$. To avoid unnecessary formalism, we do not introduce these operators explicitly. But note that, strictly, (\ref{fullIIone}) is already a linear combination of components from two regions related by $\Refl^\lii$.} It is also necessary to include a component in which the particles are at the same site, combining to form the composite excitation $\psi^4$:
\beaa \ket{\psi^4(y_1,y_2)}^\lii &=& \sum_{k=1}^\KI \ket{\psi_1^3 \dots \psi_k^4 \dots \psi_\KI^3}
      \prod_{\ell=1}^{k-1} \SIIcI(y_1;\x_\ell) \fL(y_1;\x_k,\eta_k)
      \\ &&\qquad\qquad\qquad\qquad\,\,\,\,\,\times
      \prod_{\ell=1}^{k-1} \SIIcI(y_2;\x_\ell) \fL(y_2;\x_k,\eta_k)\nn \ff(y_1,y_1;\x_k,\eta_k,\phasef_k) .\eeaa
The unknown functions are $M,N$ and $\ff$. Let us recall how they are computed, since it is a useful warm-up for the boundary calculation below. Consider a level $I$ state of $\KI=2$ particles. The overlap of $\ket{\Psi^\lii_{(y_1,a;\, y_2,b)}}$ with $\alf(\ket{\phi_1^a\phi_2^b}^\li\pm \ket{\phi_1^b\phi_2^a}^\li)$ is
\beaa \acoeff^\pm :=
\btp \draw[dotted] (2,-2) -- (2,2);
     \draw[dotted] (1,-2) -- (1,2);
     \draw[->] (-1,-2) node[below] {$_1$} -- (1,1);
     \draw[->] ( 0,-2) node[below] {$_2$} -- (2,1);
\etp +
\btp \draw[dotted] (3,-2) -- (3,2);
     \draw[dotted]     (2,-2) -- (2,2);
     \draw[->] (0,-2) node[below] {$_1$} -- (3,1);
     \draw[->] (1,-2) node[below] {$_2$} -- (2,1);
\etp &=& \fL(y_1; \x_1,\eta_1) \SIIcI(y_2;\x_1) \fL(y_2;\x_2,\eta_2) \\
 &&{}  + \fL(y_2;\x_1,\eta_1) \SIIcI(y_1;\x_1) \fL(y_1;\x_2,\eta_2)
         (M(y_1,y_2) \pm N(y_1,y_2))\nn
\eeaa
while its overlap with $\alf(\ket{\psi_1^3\psi_2^4}^\li\pm \ket{\psi_1^4\psi_2^3}^\li)$ is
\beaa \fcoeff^\pm :=\pm
\btp \draw[dotted] (3,-2) -- (3,2);
     \draw[dotted]     (2,-2) -- (2,2);
     \filldraw[fill=black] (2,1) circle (1.5mm);
     \draw[-] (0,-2) node[below] {$_1$} -- (2,1);
     \draw[-] (1,-2) node[below] {$_2$} -- (2,1);
\etp +
\btp \draw[dotted] (3,-2) -- (3,2);
     \draw[dotted]     (2,-2) -- (2,2);
     \draw[-] (0,-2) node[below] {$_1$} -- (3,1);
     \draw[-] (1,-2) node[below] {$_2$} -- (3,1);
     \filldraw[fill=black] (3,1) circle (1.5mm);
\etp
 &=& \pm \fL(y_1;\x_1,\eta_1)  \fL(y_2;\x_1,\eta_1) \ff(y_1,y_2;\x_1,\eta_1,\phasef) \\
&&{}+  \SIIcI(y_1;\x_1) \fL(y_1;\x_2,\eta_2) \SIIcI(y_2;\x_1) \fL(y_2;\x_2,\eta_2) \nn\\
&&{}\;\;\  \times \ff(y_1,y_2;\x_2,\eta_2,\phasef e^{ip_1}).\nn
\eeaa
In terms of these shorthands, $\acoeff$ and $\fcoeff$, and the coefficient functions $A,\dots,L$ of the level $\li$ scattering matrix in table \ref{fig:bulkS}, the consistency condition
\be \Smat^\li_{12} \ket{\Psi^\lii_{(y_1,a;\, y_2,b)}} = \ket{\Psi^\lii_{(y_1,a;\, y_2,b)}}_{\sigma_{12}} \SIcI \ee
reads as follows, component by component:
\beaa \ket{\phi_1^{(a}\phi_2^{b)}}_{\sigma_{12}}^\li : &\qquad&
       A \acoeff^+ = D \acoeff^+_{\sigma_{12}} \\
   \ket{\phi_1^{[a}\phi_2^{b]}}_{\sigma_{12}}^\li : &\qquad &
       B \acoeff^-  + F\fcoeff^-  = D \acoeff^-_{\sigma_{12}} \\
   \ket{\psi_1^{[3}\psi_2^{4]}}_{\sigma_{12}}^\li : &\qquad &
       C \acoeff^-  + E\fcoeff^- = D \fcoeff^-_{\sigma_{12}} \\
   \ket{\psi_1^{(3}\psi_2^{4)}}_{\sigma_{12}}^\li  : &\qquad &
       D \fcoeff^+ = D \fcoeff^+_{\sigma_{12}}.\eeaa
The first of these equations yields $M(y_1,y_2)=-1-N(y_1,y_2)$. By considering the second or third, one notices that the phase $\phasef$ dependence of $\ff$ must be $\sim 1/\phasef$. In the fourth equation, recall how these phases transform: on the right-hand side of the equation it is $\ff(y_1,y_2;\x_2,\etap_2,\phasef)$ and $\ff(y_1,y_2;\x_1,\etap_1,\phasef e^{ip_2})$ that appear.  The equation is then separable, with solutions
\be \ff(y_1,y_2; \x,\eta,\zeta) = \frac{\eta^2}{\zeta} \frac{(\xp\xm -y_1y_2)}{\xp(\xp-\xm)} \tilde\ff(y_1,y_2)\ee
for any function $\tilde\ff(y_1,y_2)$.
Finally both remaining unknowns $\tilde\ff(y_1,y_2)$ and $N(y_1,y_2)$ are fixed by the second and third equations. At this step for the first time it is necessary to make use of the mass-shell condition in (\ref{sp}). The solution is
\be M(y_1,y_2) =    \frac{\frac i g} {v_1 - v_2 - \frac i g},\qquad
N(y_1,y_2) = - \frac{v_1 - v_2}{v_1 - v_2 - \frac i g}, \ee
\be \tilde\ff(y_1,y_2) = -\frac{\frac{i}{y_1}-\frac{i}{y_2}}{v_1 - v_2 - \frac i g },\qquad
\text{where}\quad v_i=y_i + \frac{1}{y_i}.\label{SIIsoln}\ee

The calculation above was for level $\lii$ excitations whose tails both trailed to the left. But the same result holds when (\ref{bothleft}) is replaced with an asymptotic piece in which either or both tails run to the right, in the sense of (\ref{righttail}). The calculation is essentially the same: we omit the details but, for example, the pictures in the case with the tail of $y_1$ trailing to the left and the tail of $y_2$ trailing to the right are
\be \acoeff_\text{left-right} \sim
\btp \draw[dotted] (2,-2) -- (2,2);
     \draw[dotted] (1,-2) -- (1,2);
     \draw[->] ( 0,-2) node[below] {$_1$} -- (1,1);
     \draw[->] ( 3,-2) node[below] {$_2$} -- (2,1);
\etp +
\btp \draw[dotted] (3,-2) -- (3,2);
     \draw[dotted] (2,-2) -- (2,2);
     \draw[->] (1,-2) node[below] {$_1$} -- (3,1);
     \draw[->] (4,-2) node[below] {$_2$} -- (2,1);
\etp \qquad
\fcoeff_\text{left-right} \sim
\btp \draw[dotted] (3,-2) -- (3,2);
     \draw[dotted] (2,-2) -- (2,2);
     \filldraw[fill=black] (2,1) circle (1.5mm);
     \draw[-] (1,-2) node[below] {$_1$} -- (2,1);
     \draw[-] (4,-2) node[below] {$_2$} -- (2,1);
\etp +
\btp \draw[dotted] (3,-2) -- (3,2);
     \draw[dotted]     (2,-2) -- (2,2);
     \draw[-] (1,-2) node[below] {$_1$} -- (3,1);
     \draw[-] (4,-2) node[below] {$_2$} -- (3,1);
     \filldraw[fill=black] (3,1) circle (1.5mm);
\etp .\ee

\paragraph{Two particles: boundary.}We are ready to re-introduce the boundary, this time for states of $\KII=2$ excitations. We have almost all the needed ingredients: we know how to scatter two level $\lii$ excitations in the bulk (\ref{IIbulk}) and how to scatter a level $\lii$ excitation from the boundary (\ref{fullIIone}). So, starting from a component (\ref{bothleft}) in the region in which the level $\lii$ particles are ordered $1,2$ and are both heading towards the boundary, we can certainly construct a state $\big|\Psi_{(y_1,a; y_2,b)}^\lii\big\rangle$  which solves the compatibility condition \begin{enumerate} \item everywhere in the bulk, and \item at the boundary whenever only one level $\lii$ excitation lies on or next to the boundary. \end{enumerate} The one remaining case is when \emph{both} level $\lii$ excitations are on sites  $\{\KI,R\}$. Correspondingly, there is one new term we can introduce in the state vector, namely the term in which two level $\lii$ particles coincide \emph{at the boundary}:
\beaa  \btp \begin{scope}[scale=.85]\draw[very thick] (3,-2) -- (3,2);
          \draw[dotted] (2,-2) -- (2,2);
          \draw[dotted] (1,-2) -- (1,2);
          \draw[-] (-1,-2) node[below] {$_1$} -- (3,1);
          \draw[-] (0,-2) node[below] {$_2$} -- (3,1);
     \filldraw[fill=white] (3,1) circle (1.5mm);
      \end{scope}
\etp &=&
     \ket{\psi_1^3\psi_2^3 \dots \dots\psi_\KI^3\psi_R^4}
           \prod_{\ell=1}^{\KI} \SIIcI(y_1;\x_\ell) \fB(y_1;x_B,\eta_B)\\
&&\qquad\qquad\qquad\qquad \,\,\times
 \prod_{\ell=1}^{\KI} \SIIcI(y_2;\x_\ell) \fB(y_2;x_B,\eta_B)
                                      \ffB(y_1,y_2;x_B,\eta_B,f_B).\nn\eeaa
To fix the final unknown function $\ffB$ it suffices to consider a state of $\KII=2$ level $\lii$ particles on a background level $\li$ chain with only $\KI=1$ bulk sites (plus the boundary site). The compatibility condition we have to solve is
\be \Refl^\li \ket{\Psi_{(y_1,a; y_2,b)}^\lii} = \ket{\Psi_{(y_1,a; y_2,b)}^\lii}_{\tau}\, \RI
                                   = \ket{\Psi_{(y_1,a; y_2,b)}^\lii}_{\tau} .\label{cc3}\ee
Now the overlap of the full state vector  $\big|\Psi_{(y_1,a; y_2,b)}^\lii\big\rangle$ with $\alf(\ket{\phi^a\phi_R^b}^\li\pm \ket{\phi^b\phi_R^a}^\li)$ is
\beaa \acoeff_B^\pm &:=&
  \btp\begin{scope}[scale=.9]\draw[very thick] (3,-2) -- (3,2);
     \draw[dotted]     (2,-2) -- (2,2);
     \draw[->] (0,-2) node[below] {$_1$} -- (2,1);
     \draw[->] (1,-2) node[below] {$_2$} -- (3,1);
     \end{scope}
\etp \,\,\,+
\btp \begin{scope}[scale=.9]\draw[very thick] (3,-2) -- (3,2);
     \draw[dotted]     (2,-2) -- (2,2);
     \draw[->] (0,-2) node[below] {$_1$} -- (3,1);
     \draw[->] (1,-2) node[below] {$_2$} -- (2,1);
      \end{scope}
\etp \,\,\,+
\btp \begin{scope}[scale=.9]\draw[very thick] (3,-2) -- (3,2);
     \draw[dotted]     (2,-2) -- (2,2);
     \draw[->] (0,-2) node[below] {$_1$} -- (3,1);
     \draw[->] (1,-2) node[below] {$_2$} -- (3,-.5) -- (2,1);
      \end{scope}
\etp \,\,\,+
\btp \begin{scope}[scale=.9]\draw[very thick] (3,-2) -- (3,2);
     \draw[dotted]     (2,-2) -- (2,2);
     \draw[->] (0,-2) node[below] {$_1$} -- (3,-.5) -- (2,1);
     \draw[->] (1,-2) node[below] {$_2$} -- (3,1);
      \end{scope}
\etp \label{arpic}\\
&=&    \fL(y_1;\x,\eta) \SIIcI(y_2;\x) \fB(y_2;x_B,\eta_B) \nn\\
 &&{}+  \SIIcI(y_1;\x) \fB(y_1;x_B,\eta_B) \fL(y_2;\x,\eta)
         \left[M(y_1,y_2) \pm N(y_1,y_2)\right]\nn\\
&&{}+ \SIIcI(y_1;\x) \fB(y_1;x_B,\eta_B) \SIIcI(y_2;\x) \RII(y_2;x_B) \fR(\x,\eta;-y_2)\nn\\
&&\qquad \times           \left[M(y_1,-y_2) \pm N(y_1,-y_2) \right] \nn\\
&&{}+ \SIIcI(y_1,\x) \RII(y_1;x_B) \fR(\x,\eta;-y_1) \SIIcI(y_2;\x,\eta) \fB(y_2;x_B,\eta_B)\nn\\
&&\qquad \times  \big[M(y_1,y_2)M(y_2,-y_1)\pm N(y_1,y_2)M(y_2,-y_1) \nn \\
&&\qquad\quad{}   \pm M(y_1,y_2)N(y_2,-y_1)+N(y_1,y_2)N(y_2,-y_1)\big]\eeaa
while its overlap with $\alf(\ket{\psi^3\psi_R^4}^\li\pm \ket{\psi^4\psi_R^3}^\li)$ is
\beaa\fcoeff_B^\pm &:=&
\btp \begin{scope}[scale=.9]\draw[very thick] (3,-2) -- (3,2);
     \draw[dotted]     (2,-2) -- (2,2);
     \draw[-] (0,-2) node[below] {$_1$} -- (3,1);
     \draw[-] (1,-2) node[below] {$_2$} -- (3,1);
     \filldraw[fill=white] (3,1) circle (1.5mm);
\end{scope}
\etp \,\,\, \pm
\btp \begin{scope}[scale=.9]\draw[very thick] (3,-2) -- (3,2);
     \draw[dotted]     (2,-2) -- (2,2);
     \filldraw[fill=black] (2,1) circle (1.5mm);
     \draw[-] (0,-2) node[below] {$_1$} -- (2,1);
     \draw[-] (1,-2) node[below] {$_2$} -- (2,1);
     \end{scope}
\etp\,\,\, \pm
\btp \begin{scope}[scale=.9]\draw[very thick] (3,-2) -- (3,2);
     \draw[dotted]     (2,-2) -- (2,2);
     \filldraw[fill=black] (2,1) circle (1.5mm);
     \draw[-] (0,-2) node[below] {$_1$} -- (3,0) -- (2,1);
     \draw[-] (1,-2) node[below] {$_2$} -- (2,1);
\end{scope}
\etp\,\,\, \pm
\btp \begin{scope}[scale=.9]\draw[very thick] (3,-2) -- (3,2);
     \draw[dotted]     (2,-2) -- (2,2);
     \filldraw[fill=black] (2,1) circle (1.5mm);
     \draw[-] (0,-2) node[below] {$_1$} -- (2,1);
     \draw[-] (1,-2) node[below] {$_2$} -- (3,0) -- (2,1);
     \end{scope}
\etp\,\,\, \pm
\btp \begin{scope}[scale=.9]\draw[very thick] (3,-2) -- (3,2);
     \draw[dotted]     (2,-2) -- (2,2);
     \filldraw[fill=black] (2,1) circle (1.5mm);
     \draw[-] (0,-2) node[below] {$_1$} -- (3,0.5) -- (2,1);
     \draw[-] (1,-2) node[below] {$_2$} -- (3,-1) -- (2,1);
\end{scope}
\etp \label{frpic}\\
&=& \SIIcI(y_1;\x) \fB(y_1;x_B,\eta_B) \SIIcI(y_2;\x) \fB(y_2;x_B,\eta_B)
                  \ffB(y_1,y_2;x_B,\eta_B,\phasef e^{ip} ) \nn\\
&&{}\pm \fL(y_1;\x,\eta) \fL(y_2;\x,\eta) \ff(y_1,y_2;\x,\eta,\phasef) \nn\\
&&{}\pm \SIIcI(y_1;\x) \RII(y_1;x_B) \fR(\x,\eta;-y_1) \fL(y_2;\x,\eta)
    \ff(y_2,-y_1;\x,\eta,\phasef) \nn\\
&&{} \quad\times    \left[ M(y_1,y_2) - N(y_1,y_2)\right] \nn\\
&&{}\pm \fL(y_1;\x,\eta) \SIIcI(y_2;\x) \RII(y_2;x_B) \fR(\x,\eta;-y_2)
    \ff(y_1,-y_2;\x,\eta,\phasef) \nn\\
&&{}\pm  \SIIcI(y_1;\x) \RII(y_1;x_B) \fR(\x,\eta;-y_1)
         \SIIcI(y_2;\x) \RII(y_2;x_B) \fR(\x,\eta;-y_2)\nn\\
&&{} \quad\times\ff(-y_2,-y_1;\x,\eta,\phasef) \left[ M(y_1,-y_2) - N(y_1,-y_2) \right].\eeaa
Here it is necessary to think rather carefully about which terms should be included. Let us comment on this.

Recall the structure of a coordinate Bethe ansatz: there is always a component in the state vector for each region, i.e. each Weyl chamber of the relevant reflection group, here $BC_2$. Neighbouring Weyl chambers meet at one of the mirrors, where a compatibility condition must be met.
In the present case it was necessary to include additional components  (the $\fB$ and $\ff$ terms, respectively) associated to boundaries between neighbouring regions, which are subsets of the $\tau$ and $\sigma_{12}$ mirrors themselves. And finally, the $\ffB$ term is associated to the \emph{intersection} of the $\sigma_{12}$ mirror with the $\tau$ mirror.

With this structure in mind, it is possible systematically to list all the ways in which the two particles can end up next to and on the boundary. One finds that only those processes pictured in (\ref{arpic}) are valid. For example, one might be tempted to include
\btp \begin{scope}[scale=.8]\draw[very thick] (3,-2) -- (3,2);
     \draw[dotted]     (2,-2) -- (2,2);
     \draw[->] (0,-2) node[below] {$_1$} -- (3,.5) -- (2,1);
     \draw[->] (1,-2) node[below] {$_2$} -- (3,-.5);
\end{scope}s\etp\, and
\btp \begin{scope}[scale=.8]\draw[very thick] (3,-2) -- (3,2);
     \draw[dotted]     (2,-2) -- (2,2);
     \draw[->] (0,-2) node[below] {$_1$} -- (3,-.5);
     \draw[->] (1,-2) node[below] {$_2$} -- (3,.5) -- (2,1);
     \end{scope}
\etp\,. But in the first of these, $y_2$ is initially the particle closest to the boundary, so $y_1$ cannot in fact reach the boundary and reflect until it has intersected the path of $y_2$. And likewise in the second diagram after the first scattering of $y_1$ with $y_2$.

Similarly it is possible to list all the ways in which \emph{both} particles can end up at the site next to boundary, and find the final four diagrams in (\ref{frpic}). In doing so, one should consider also the process
\beaa \smash{
\btp \draw[very thick] (3,-2) -- (3,2);
     \draw[dotted]     (2,-2) -- (2,2);
     \filldraw[fill=black] (2,1) circle (1.5mm);
     \draw[-] (0,-2) node[below] {$_1$} -- (3,-1) -- (2,1);
     \draw[-] (1,-2) node[below] {$_2$} -- (3,.5) -- (2,1);
\etp}&=&{}\pm  \SIIcI(y_1;\x) \RII(y_1;x_B) \fR(\x,\eta;-y_1)
         \SIIcI(y_2;\x) \RII(y_2;x_B) \fR(\x,\eta;-y_2)
     \nn\\
&& \times \ff(-y_1,-y_2;\x,\eta,\phasef)  \big[M(y_1,y_2)M(y_2,-y_1)-N(y_1,y_2)M(y_2,-y_1) \nn \\
&&\qquad\qquad\qquad\qquad\qquad{}   - M(y_1,y_2)N(y_2,-y_1)+N(y_1,y_2)N(y_2,-y_1)\big].\eeaa
This is a valid sequence of scattering events. But observe that it produces the term in the ansatz associated to the boundary between the following two regions: both particles outgoing, ordered $y_1$, $y_2$; and both particles outgoing, ordered $y_2$, $y_1$. We have already included a term associated to this boundary: it is the final term in (\ref{frpic}). And indeed these terms turn out to be equal, as they must be. So one should include one or other but not both.

Having found the overlap functions $\acoeff_R^\pm$ and $\fcoeff_R^\pm$ for the boundary, we can plug them in to the consistency condition, which is, once more component by component,
\beaa \ket{\phi_1^{(a}\phi_2^{b)}}_{\tau}^\li : &\qquad&
       A \acoeff_R^+ = D (\acoeff_R^+)_{\tau} \\
   \ket{\phi_1^{[a}\phi_2^{b]}}_{\tau}^\li : &\qquad &
       B \acoeff_R^-  + F\fcoeff_R^-  = D (\acoeff_R^-)_{\tau} \\
   \ket{\psi_1^{[3}\psi_2^{4]}}_{\tau}^\li : &\qquad &
       C \acoeff_R^-  + E\fcoeff_R^- = D (\fcoeff_R^-)_{\tau} \\
   \ket{\psi_1^{(3}\psi_2^{4)}}_{\tau}^\li  : &\qquad &
       D \fcoeff_R^+ = D (\fcoeff_R^+)_{\tau}.\eeaa
The first of these does not include the new unknown $\ffB$, and is satisfied upon inserting the level $\lii$ scattering matrix in (\ref{SIIsoln}). On inspecting the second or third one sees that $\ffB$ must go like $\eta_B^2/\phasef$. One then looks for a solution to the forth equation of this form, and finds
\be \ffB(y_1,y_2;x_B,\eta_B,\phasef)=  \frac{i \eta_B^2}{\phasef}
\frac{(\frac{1}{y_1}-\frac{1}{y_2})(\frac{1}{y_1}+\frac{1}{y_2})(1 - \frac{iy_1y_2}{x_B})(1+\frac{iy_1y_2}{x_B})}{(v_1-v_2-\frac i g)(v_1+v_2-\frac i g)}.\ee
Given the mass shell conditions, we have verified that all four equations are then satisfied.

At this stage we have solved for all the functions that appear in the level $\lii$ ansatz, and  demonstrated that it works for states of $\KII=2$ particles. No new types of \emph{terms} arise for states of $\KII>2$ particles and, since the original problem is solvable in the sense discussed after (\ref{BCK}), one can be confident that the ansatz continues to work. This is, admittedly, not quite manifest because there are superficially new types of compatibility \emph{conditions} to check when $\KII>2$. But these are not exclusive to our present boundary case: even in the bulk one sees for the first time $\psi^4\phi^a$ appearing as neighbouring spins in the level $\lii$ chain.

\subsubsection*{Level III}
Finally we come to level $\liii$ of the nesting. The goal is much as it was in going from level $\li$ to $\lii$: we know that a component
\be \ket{\phi^a(y_1)\phi^b(y_2) \dots \phi^z(y_\KII)}^\lii \label{liibasis}\ee
of a level $\lii$ state in any one region can be uniquely completed, by including the terms for all other regions and the additional terms for boundaries of regions, to a Bethe state $\big|\Psi^\lii_{(y_1,\,a;\,y_2,\,b;\,\dots;\,y_\KII,\,z)}\big\rangle$ obeying (\ref{BetheStateI}). But the $su(2)$ indices $a,b,\dots, z$ will in general be transformed non-trivially by these level $\lii$ scattering processes. We want to identify those linear combinations of states (\ref{liibasis}) on which the level $\lii$ scattering operators act diagonally.

Let the level $\liii$ vacuum be
\be \vac^\liii = \ket{\phi^1(y_1)\phi^1(y_2) \dots \phi^1(y_\KII)}^\lii.\ee
For single particles in the bulk we again make a spin-wave ansatz,
\be \ket{\phi^2(w)}^{\liii} = \sum_{k=1}^\KII \ket{\phi^1(y_1)\dots \phi^2(y_k)\dots \phi^1(y_\KII)}^\lii \prod_{\ell=1}^{k-1} \SIIIcII(w,y_\ell) \hL(w,y_k),\ee
and find that the bulk compatibility condition (cf. \ref{IIbulk} and \ref{SIIsoln}) is solved by
\be \hL(w,y) = \frac{\frac{i}{2g}}{w-v -\frac{i}{2g}},\qquad
\SIIIcII(w,y)= \frac{w-v+\frac{i}{2g}}{w-v-\frac{i}{2g}}.\ee
Then, defining $\SIIcIII$ and $\hR$ just as at level $\lii$, cf. (\ref{sicii}-\ref{fr}), we can make an ansatz for a single particle in the presence of a right boundary as in (\ref{fullIIone}), except that there is no distinguished boundary site for the level $\liii$ chain and so no boundary term in the ansatz. Finally, after also solving for $\SIIIcIII$ component of the diagonalized scattering matrix, one has
\be \RIII(w) = -1,\qquad \SIIIcIII(w_1,w_2) = \frac{w_1-w_2-\frac i g}{w_1 - w_2 + \frac i g}\,.\ee

\subsection{Bethe Equations}\label{Beqns}

The nested coordinate Bethe ansatz above was for the semi-infinite system with a right boundary. Let us now add the left boundary, so placing the system on a finite interval. Then the Bethe equations are the quantization conditions obtained as follows: starting from any given component of the state vector, consider picking up a particle (belonging to any level, $\li$, $\lii$ or $\liii$, of the nesting), moving it through all the particles lying its right, reflecting it from the right boundary, moving it back again through all the particles, reflecting it from the left boundary, and finally moving it through all the particles that were originally its left:
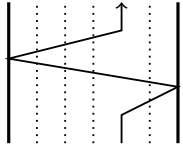
\be\nn\btp\begin{scope}[scale = 1.5]
 \draw[very thick] (3,-2) -- ++(0,5);
\draw[very thick] (-3,-2) -- ++(0,5);
        \draw[dotted] (2,-2) -- ++(0,5);
        \draw[dotted] (0,-2) -- ++(0,5);
        \draw[dotted] (-1,-2) -- ++(0,5);
        \draw[dotted] (-2,-2) -- ++(0,5);
        \draw[->] (1,-2) -- (1,-1) -- (3,0) -- (-3,1) -- (1,2) -- (1,3);
\end{scope}\etp\ee
Since this sequence of operations returns all the particles to their initial positions and (quasi)rapidities, we must recover the component of the state vector we began with -- modulo, in case of the physical i.e. level $\li$ particles, a phase factor ``$e^{2ipL}$'' that comes from translating to the right a total distance $L$ with momentum $p$ and a distance $L$ to the left with reflected momentum $-p$. Here we add scare-quotes, because we need to be more precise about the meaning of the system size $L$.
Thus, the Bethe equations take the form:
\be  R_R^A(x^A_k) R_L^A(-x^A_k) \underset{(A,k)\neq (B,\ell)}{\prod_{B=\li}^\liii \prod_{\ell = 1}^{K^B}} S^{A,B}(x^A_k,x^B_\ell) S^{B,A}(x^B_\ell,-x^A_k) =
\left\{\begin{array}{cl}
 \left(\frac{\xp_k}{\xm_k}\right)^{-2L} & {\rm for\ } A=\li
\\
1 & {\rm for\ } A=\lii,\liii.
\end{array}\right.
\ee
Here $x^A$ denotes the relevant rapidity variable for a particle at level $A\in \{\li,\lii,\liii\}$, and $-x^A$ the reflected rapidity: thus in particular $x^\li = \xpm$, $-x^\li = \mxmp$.  We can use parity symmetry to write $R^A_L(-x^A) = R^A_R(x^A)$ and
$S^{B,A}(x^B_\ell,-x^A_k)=S^{A,B}(x^A_k,-x^B_\ell)$ in the equations.
Explicitly then, the Bethe equations for the $su(2|2)$ scattering theory with ``$Z=0$'' boundaries are
\beaa \label{seve}
1 &=& R_{0{\rm R}}(\xpm_k)^2 \left(\frac{\xp_k}{\xm_k}\right)^{2L}
         \ \prod_{\ell\neq k}^\KI S_0(\xpm_k,\xpm_\ell) S_0(\xpm_k,\mxmp_\ell)
         \prod_{\ell=1}^\KII \frac{y_\ell-\xm_k}{y_\ell-\xp_k} \frac{y_\ell+\xm_k}{y_\ell+\xp_k}\\
      1 &=& \left(\frac{y_k- x_B}{y_k+x_B}\right)^2
         \prod_{\ell=1}^\KI  \frac{y_k-\xp_\ell}{y_k-\xm_\ell} \frac{y_k+\xm_\ell}{y_k+\xp_\ell}
         \prod_{\ell=1}^\KIII \frac{w_\ell-v_k -\frac{i}{2g}}{w_\ell-v_k+\frac{i}{2g}}
                              \frac{w_\ell+v_k +\frac{i}{2g}}{w_\ell+v_k-\frac{i}{2g}} \\
      1 &=&
         \prod_{\ell=1}^\KII \frac{w_k-v_\ell +\frac{i}{2g}}{w_k-v_\ell-\frac{i}{2g}}
                               \frac{w_k+v_\ell -\frac{i}{2g}}{w_k+v_\ell+\frac{i}{2g}}
         \prod_{\ell\neq k}^\KIII \frac{w_k - w_\ell - \frac i g}{w_k - w_\ell + \frac i g}
                                 \frac{w_k + w_\ell - \frac i g}{w_k + w_\ell + \frac i g}.
                                 \eeaa

To be more precise about the meaning of $L$, we can consider the equations in the weak coupling limit, where they should be those of an open spin chain. It suffices to consider the case of a single level I excitation. We have to specify how our definition of the reflection factor relates the in-going and out-going spin-waves. We do that explicitly in the appendix.

In the conventions we are following, the $R_{0}$ appearing in the Bethe equations would be the overall scalar factor of Hofman and Maldacena \cite{HMopen}, times the corresponding dressing factors to satisfy the boundary crossing symmetry condition \cite{ChCo,ABR},
\ba
R_{0{\rm L}}^2 = -\frac{(x^-)^2(x_B-x^-)(x_B+x^-)(x_B+\tfrac{1}{x^+})(x_B +\frac{1}{x^-})}{(x^+)^2(x_B-x^+)(x_B+x^+)(x_B-\tfrac{1}{x^-})(x_B-\frac{1}{ x^+})}
 \, \sigma(x,-x) \sigma^2(x,\pm x_B).
\ea

Let us take for instance (\ref{seve}) for $\KI=1$ and $\KIIn\alpha=0$. This should reproduce the Bethe equation for single particle in the
$sl(2)$ sector which reads (\ref{sleq})
\be
R_{0{\rm L}}^4(x) = \left(\frac{\xp}{\xm}\right)^{2(K^0-1)},
\ee
where $K^0$ is the number of sites in the underlying spin-chain (including the boundary sites). This implies that we have to take
$L=K^0-1$ in (\ref{seve}).

At this point we should recall that, for the operators  we are considering, the symmetry is actually $su(2|2)^2$ and that the excitations are in bifundamental representations. This simply means there are two kinds of level $\lii$ and level $\liii$ particles, indexed by $\alpha=1,2$. So the full Bethe equations read
\beaa
\label{baev3I}
  1 \!\!&=&\!\!\!  R_{0{\rm R}}(\xpm_k)^4 \left(\frac{\xp_k}{\xm_k}\right)^{\!\!2(K^0-1)}
         \!\prod_{\ell\neq k}^\KI S_0(\xpm_k,\xpm_\ell)^2 S_0(\xpm_k,\mxmp_\ell)^2
         \!\prod_{\alpha=1}^2\!\prod_{\ell=1}^{K_{(\alpha)}^\lii}\! \frac{y_{\ell}^{(\alpha)}\!-\xm_k}{y_{\ell}^{(\alpha)}\!-\xp_k} \;\frac{y_{\ell}^{(\alpha)}\!+\xm_k}{y_{\ell}^{(\alpha)}\!+\xp_k}\\
\label{baev3II}
      1 \!\!&=&\! \! \!\left(\frac{y_k^{(\alpha)}- x_B}{y_k^{(\alpha)}+x_B}\right)^2\!
         \prod_{\ell=1}^\KI  \frac{y_{k}^{(\alpha)}-\xp_\ell}{y_{k}^{(\alpha)}-\xm_\ell} \;\;\frac{y_{k}^{(\alpha)}+\xm_\ell}{y_{k}^{(\alpha)}+\xp_\ell}
         \prod_{\ell=1}^\KIIIA \frac{w_\ell^{(\alpha)}-v_k^{(\alpha)} -\frac{i}{2g}}{w_\ell^{(\alpha)}-v_k^{(\alpha)}+\frac{i}{2g}}
                             \;\; \frac{w_\ell^{(\alpha)}+v_k^{(\alpha)} +\frac{i}{2g}}{w_\ell^{(\alpha)}+v_k^{(\alpha)}-\frac{i}{2g}} \\
\label{baev3III}
      1 \!\!&=&\!
         \prod_{\ell=1}^\KIIA \frac{w_k^{(\alpha)}-v_\ell^{(\alpha)} +\frac{i}{2g}}{w_k^{(\alpha)}-v_\ell^{(\alpha)}-\frac{i}{2g}}\;\;
                               \frac{w_k^{(\alpha)}+v_\ell^{(\alpha)} -\frac{i}{2g}}{w_k^{(\alpha)}+v_\ell^{(\alpha)}+\frac{i}{2g}}
         \prod_{\ell\neq k}^\KIIIA \frac{w_k^{(\alpha)} - w_\ell^{(\alpha)} - \frac i g}{w_k^{(\alpha)} - w_\ell^{(\alpha)} + \frac i g}
                                \;\; \frac{w_k^{(\alpha)} + w_\ell^{(\alpha)} - \frac i g}{w_k^{(\alpha)} + w_\ell^{(\alpha)} + \frac i g}.\eeaa

But $K^0$ is not a good quantum number, because,  beyond one-loop, the length of the chain can vary under mixing \cite{B2003}. We would therefore like to eliminate it, in favour of the R-charge $J=J_{56}$ (which, being the Noether charge of a symmetry of the quantum theory, is certainly a good quantum number).
More precisely, we will eliminate $K^0$ in favour of $J\equiv J_{\rm string}=J_{\rm total}-N+1$, where  $J_{\rm total}$ is the total R-charge $J_{56}$ in the operators (\ref{zzeroop}) we are considering.

We can translate states $\chi^{\mathrm a,\dot{\mathrm a}}$ into fields of the $\mathcal N=4$ action
and specify how much they contribute to $J$ and $K^A$.
\be \begin{array}{l|c|cccc} \label{N4fields}
                            & J  & K^0 & \KI & \KIIn1 & \KIIn2      \\\hline
\chi^{1,\dot 1}\sim\Phi     & 0  &  1  &   1 &    1     &    1      \\
\chi^{3,\dot 1}\sim\Psi     &\alf&  1  &   1 &    1     &    0      \\
\chi^{1,\dot 3}\sim\bar\Psi &\alf&  1  &   1 &    0     &    1      \\
\chi^{3,\dot 3}\sim D_\mu Z & 1  &  1  &   1 &    0     &    0      \\
Z                           & 1  &  1  &   0 &    0     &    0      \end{array}
\ee
In each case $J=K^0-\alf \KIIn1 -\alf \KIIn2$. Thus, the total contribution to $J$
is
\be J=  K^0-\alf \KIIn1 -\alf \KIIn2 \ee
and the Bethe equations can be re-written as
{\small\beaa
1   \!\!\!\!&=&\!\!\!\!
\label{baev3I2}
 R_0(\xpm_k)^4 \left(\tfrac{\xp_k}{\xm_k}\right)^{\!\!2(J-1)}
       \!  \prod_{\ell\neq k}^\KI S_0(\xpm_k,\xpm_\ell)^2 S_0(\xpm_k,\mxmp_\ell)^2
\!\prod_{\alpha=1}^2\!\prod_{\ell=1}^{K_{(\alpha)}^\lii}\!\! \sqrt{\tfrac{\xp_k}{\xm_k}}\frac{y_{\ell}^{(\alpha)}\!-\xm_k}{y_{\ell}^{(\alpha)}\!-\xp_k} \sqrt{\tfrac{\xp_k}{\xm_k}}\frac{y_{\ell}^{(\alpha)}\!+\xm_k}{y_{\ell}^{(\alpha)}\!+\xp_k}\\
\label{baev3II2}
      1 \!\!\! \!&=&\! \! \!\!\!\! \left(\frac{y_k^{(\alpha)}\!- x_B}{y_k^{(\alpha)}\!+x_B}\right)^2\!
         \prod_{\ell=1}^\KI  \sqrt{\tfrac{\xp_\ell}{\xm_\ell}} \frac{y_{k}^{(\alpha)}\!-\xp_\ell}{y_{k}^{(\alpha)}\!-\xm_\ell} \sqrt{\tfrac{\xm_\ell}{\xp_\ell}}    \frac{y_{k}^{(\alpha)}\!+\xm_\ell}{y_{k}^{(\alpha)}\!+\xp_\ell}
         \prod_{\ell=1}^\KIIIA \frac{w_\ell^{(\alpha)}\!-v_k^{(\alpha)}\! -\frac{i}{2g}}{w_\ell^{(\alpha)}\!-v_k^{(\alpha)}\!+\frac{i}{2g}}
                             \;\; \frac{w_\ell^{(\alpha)}\!+v_k^{(\alpha)}\! +\frac{i}{2g}}{w_\ell^{(\alpha)}\!+v_k^{(\alpha)}\!-\frac{i}{2g}} \\
\label{baev3III2}
      1 \!\!\!\!&=&\!
         \prod_{\ell=1}^\KIIA \frac{w_k^{(\alpha)}\!-v_\ell^{(\alpha)} \!+\frac{i}{2g}}{w_k^{(\alpha)}\!-v_\ell^{(\alpha)}\!-\frac{i}{2g}}\;\;
                               \frac{w_k^{(\alpha)}\!+v_\ell^{(\alpha)}\! -\frac{i}{2g}}{w_k^{(\alpha)}\!+v_\ell^{(\alpha)}\!+\frac{i}{2g}}
         \prod_{\ell\neq k}^\KIIIA \frac{w_k^{(\alpha)}\! - w_\ell^{(\alpha)} \!- \frac i g}{w_k^{(\alpha)}\! - w_\ell^{(\alpha)} \!+ \frac i g}
                                \;\; \frac{w_k^{(\alpha)}\! + w_\ell^{(\alpha)}\! - \frac i g}{w_k^{(\alpha)}\! + w_\ell^{(\alpha)}\! + \frac i g}.\eeaa}

\paragraph{Vacuum 1} Had we chosen $1$ as the vacuum orientation throughout rather than $3$, we would have $J=K^0-\KI-2+\alf \KIIn1 +\alf \KIIn2$, and would have obtained, by arguments paralleling those above, the Bethe equations in the following form:
{\small\beaa\label{baev1I}
  1 \!\!\!\!&=&\!\!\!  R_0(\xpm_k)^4  \left(\frac{\xp_k}{\xm_k}\right)^{2(J-1)}\! \left(\left(\tfrac{\xp_k}{\xm_k}\right)^{3/2} \frac{x_k^-(x_k^--x_B)}{x_k^+(x_k^++x_B)} \right)^4\nn\\
         && \times\prod_{\ell\neq k}^\KI S_0(\xpm_k,\xpm_\ell)^2 S_0(\xpm_k,\mxmp_\ell)^2
 \left(\sqrt{\tfrac{\xp_k\xm_\ell}{\xm_k\xp_\ell}}\frac{\xm_k-\xp_\ell}{\xp_k-\xm_\ell}\right)^2
 \left(\sqrt{\tfrac{\xp_k\xp_\ell}{\xm_k\xm_\ell}}\frac{\xm_k+\xm_\ell}{\xp_k+\xp_\ell}\right)^2
         \nn\\
         &&\times \prod_{\alpha=1}^2\prod_{\ell=1}^{K_{(\alpha)}^\lii} \sqrt{\tfrac{\xm_k}{\xp_k}}\frac{y_{\ell}^{(\alpha)}-\xp_k}{y_{\ell}^{(\alpha)}-\xm_k}    \;\sqrt{\tfrac{\xm_k}{\xp_k}}\frac{y_{\ell}^{(\alpha)}+\xp_k}{y_{\ell}^{(\alpha)}+\xm_k}
         \\
         \label{baev1II}
      1 \!\!\!\!&=&\! \! \!\!\left(\frac{y_k^{(\alpha)}\!+ x_B}{y_k^{(\alpha)}\!-x_B}\right)^2\!
         \prod_{\ell=1}^\KI  \sqrt{\tfrac{\xp_k}{\xm_k}}\frac{y_{k}^{(\alpha)}\!-\xm_\ell}{y_{k}^{(\alpha)}\!-\xp_\ell} \sqrt{\tfrac{\xm_k}{\xp_k}}\frac{y_{k}^{(\alpha)}\!+\xp_\ell}{y_{k}^{(\alpha)}\!+\xm_\ell}
         \prod_{\ell=1}^\KIIIA \frac{w_\ell^{(\alpha)}\!-v_k^{(\alpha)}\! +\frac{i}{2g}}{w_\ell^{(\alpha)}\!-v_k^{(\alpha)}\!-\frac{i}{2g}}
                             \;\; \frac{w_\ell^{(\alpha)}\!+v_k^{(\alpha)}\! -\frac{i}{2g}}{w_\ell^{(\alpha)}\!+v_k^{(\alpha)}\!+\frac{i}{2g}}
                             \\
      1 \!\!\!\!&=&\!
         \prod_{\ell=1}^\KIIA \frac{w_k^{(\alpha)}\!-v_\ell^{(\alpha)}\! -\frac{i}{2g}}{w_k^{(\alpha)}\!-v_\ell^{(\alpha)}\!+\frac{i}{2g}}\;\;
                               \frac{w_k^{(\alpha)}\!+v_\ell^{(\alpha)}\! +\frac{i}{2g}}{w_k^{(\alpha)}\!+v_\ell^{(\alpha)}\!-\frac{i}{2g}}
         \prod_{\ell\neq k}^\KIIIA \frac{w_k^{(\alpha)}\! - w_\ell^{(\alpha)}\! + \frac i g}{w_k^{(\alpha)}\! - w_\ell^{(\alpha)}\! - \frac i g}
                                \;\; \frac{w_k^{(\alpha)}\! + w_\ell^{(\alpha)}\! + \frac i g}{w_k^{(\alpha)}\! + w_\ell^{(\alpha)}\! - \frac i g}
                                \label{baev1III}
                                \eeaa}
Eq. (\ref{baev3I2})-(\ref{baev3III2}) should be, of course, equivalent to eq. (\ref{baev1I})-(\ref{baev1III}). Consider for instance an  $sl(2)$ state with a single bulk impurity, which has $\KI=1$, $\KIIn1=\KIIn2=3$ and $\KIIIn\alpha=0$. The corresponding auxiliary rapidities $y_k^{(\alpha)}$ solving eq. (\ref{baev1II})\footnote{For a single bulk impurity eq. (\ref{baev1II}) is solved by $y\to 0, \infty, \sqrt{\tfrac{x_B(x_B(\xm-\xp) +2\xp\xm )}{\xp-\xm +2x_B}} $} can be taken such that when plugged back in (\ref{baev1I}), equation (\ref{baev3I2}) for $\KI=1$ and $\KIIn1=\KIIn2=\KIIIn\alpha=0$ is recovered.

We expect the Bethe equations we presented here to be valid also for arbitrary $Q$-bound state magnons, just by using the spectral parameters satisfying the $Q$-bound state mass-shell condition. This could be shown in general by reformulating the coordinate Bethe ansatz in terms of coproducts of Yangian symmetry generators \cite{deLeeuw:2008}. For the particular case of fundamental boundary magnons and generic $Q$-bound state magnons, these Bethe equations could be obtained from a coordinate Bethe ansatz using the explicit 1-$Q$ boundary reflection matrix obtained in \cite{CY2}, much as we did here for fundamental bulk magnons.

\section{ $Z=0$ D7-brane}
\label{d7}

Another system enjoying integrable open boundary conditions is the spectral problem in the planar limit of a ${\cal N}=2$ super Yang-Mills theory, consisting of the ${\cal N}=4$ theory supplemented with a chiral hypermultiplet of fundamental matter \cite{KMMW}.

Given a {chain} of ${\cal N}=4$ fields, one can either take a trace to make a gauge invariant operator, or use the fundamental matter to contract the $SU(N)$ indices. In the planar limit, the first possibility would lead to the same spectral problem as that of ${\cal N}=4$ with periodic boundary conditions. The second possibility amounts for imposing open boundary conditions to the chain of fields. In the dual gravity description this corresponds to an open string attached to a probe D7-brane in $AdS_5\times S^5$, whose worldvolume wraps the entire $AdS_5$ and a maximal $S^3\subset S^5$.

In certain sectors, this open problem was shown to be integrable at its weak \cite{ErMa} and strong coupling \cite{Mann} limits. Moreover, we have proposed in a previous paper \cite{CY} a reflection matrix interpolating between both limits, consistent with the symmetries of the problem and satisfying the boundary Yang-Baxter equation.

As for the giant graviton cases, the relative orientation between the vacuum field $Z$ and the 3-sphere matters. If the maximal 3-sphere was defined by the intersection of $Y=0$ with $|X|^2+|Y|^2+|Z|^2=1$, only a $psu(2|1)\times psu(2|1)\subset psu(2|2)\times psu(2|2)$ would be preserved \cite{CY}. This is essentially the same problem as the $Y=0$ giant graviton. Therefore, the corresponding Bethe ansatz would be the one formulated in \cite{Galleas,Nepomechie:2009zi}.

When the maximal 3-sphere is defined through $Z=0$, we have for the reference state of the Bethe ansatz the following operator
\be
\chi^{\dot{\mathrm a}}_\LL Z^J
 \chi^{\dot{\mathrm e}}_\RR.
 \label{D7op}
 \ee
The boundary matter fields transform in a representation $(1,\boxslash)$ of $psu(2|2)\times psu(2|2)$ and only a $su(2)\times su(2) \times psu(2|2)$ symmetry is preserved by the reflection of bulk excitations.

Concerning excitations carrying fundamental dotted $psu(2|2)$ indices, their boundary reflection
\be
\Refl_R:   \boxslash_{(p,\phasef,\eta)}          \otimes \boxslash_{(\phasef e^{ip},\eta_B)} \longrightarrow
               \boxslash_{(-p,\phasef,\etap)}   \otimes \boxslash_{(\phasef e^{-ip},\etap_B)}
\ee
would be exactly the same one that is reproduced in table \ref{fig:bndryRright}. Therefore, the spin-wave functions for excitations of higher levels of nesting would also be the same ones  as those described in the section \ref{CBAgg}.

However, the problem becomes different when it comes to the undotted $psu(2|2)$ fundamental excitations, because they are reflected by a singlet boundary
\be
\Refl_R:   \boxslash_{(p,\phasef,\eta)}          \otimes 1 \longrightarrow
               \boxslash_{(-p,\phasef,\etap)}   \otimes 1.
\ee

In \cite{CY}, we have shown that the most general form of this reflection consistent with integrability is of the form
\ba
\Refl |\phi^{a}(x)\rangle \!&=&\! R_0(x)\;  \frac{\eta}{\tilde\eta}\;\frac{x_B-\xm}{x_B+\xp}|\phi^{a}(-x)\rangle,
\\
\Refl |\psi^{\alpha}(x)\rangle \!&=&\! R_0(x) |\psi^{\alpha}(-x)\rangle.
\ea

As before, a coordinate Bethe ansatz can be formulated for this case. When we introduce a boundary, the spin-wave for a single level $\lii$ excitation would be slightly simpler this time, because the excitation can not be allocated at the boundary site. The Bethe state will be the following combination of ingoing  and outgoing  spin-waves,
\beaa \ket{\Psi^\lii_{(y,a)}} &=& \dots +
   \btp \draw[very thick] (3,-2) -- (3,2);
        \draw[dotted] (2,-2) -- (2,2);
        \draw[dotted] (1,-2) -- (1,2);
        \draw[->] (0,-2)   -- (2,1);\etp \,\,\,
+  \btp \draw[very thick] (3,-2) -- (3,2);
   \draw[dotted] (2,-2) -- (2,2);
 \draw[dotted] (1,-2) -- (1,2);
        \draw[->] (0,-2)  -- (3,0) -- (2,1);\etp\,\,\,
 + \cdots \label{D7fullIIone}\\
 &=&  \sum_{k=1}^\KI \ket{\psi_1^3\psi_2^3 \dots \phi_k^a \dots\psi_\KI^3}
                 \prod_{\ell=1}^{k-1} \SIIcI(y;\x_\ell) \fL(y;{ \x_k}) \nn\\
  &&{}+ \sum_{k=1}^\KI \ket{\psi_1^3\psi_2^3 \dots \phi_k^a \dots\psi_\KI^3\psi_R^3}
                 \prod_{\ell=1}^{\KI} \SIIcI(y;\x_\ell) \RII(y;x_B) \nn\\
 &&\qquad\qquad\qquad\qquad\qquad\qquad
     \times\prod_{\ell=k+1}^{\KI} \SIcII(\x_\ell;-y) \fR({ \x_k};-y), \nn\eeaa
with only one unknown function $\RII(y;x_B)$.

The bulk compatibility condition is again satisfied by construction. For the boundary compatibility condition, it suffices to consider a state with only $\KI=1$. The compatibility condition requires then
\be
\frac{x_B-\xm}{x_B+\xp} =  \frac{\fL(y;-x) + \SIIcI(y;-x) \RII(y) \fR(-x;-y)}{\fL(y;x) + \SIIcI(y;x) \RII(y) \fR(x;-y)},
\ee
which admits the solution
\be
\RII(y;x_B)       =  \frac{y-x_B}{y+x_B}.
\label{D7fBRIIsoln}
\ee

The compatibility conditions for states with $\KII\geq2$ excitations are going to be just those of the bulk, because the boundary is a singlet and therefore level $\lii$ excitations can not be allocated there.

When moving to the next level of nesting one has got, of course, the same components $\SIIcIII$ and $\SIIIcIII$ for the diagonalized bulk scattering matrix. For the reflection of a level $\liii$ impurity we find again $\RIII(w) = -1$

To finish this section let us write down the resulting Bethe equations after the introduction of two boundaries. The first thing to note is that the $\RII(y;x_B)$ is the same (up to a sign) as in the case with the fundamental excitations in the boundary. One should also note that level II excitations of the right $psu(2|2)$ move in a chain (of level I excitations) two sites longer than the level II excitations of the left $psu(2|2)$. This so because the boundary degree of freedom have dotted (right) $psu(2|2)$ indices only. Nonetheless, this has no relevance for the Bethe equation because level II excitations do not pick up any phase when moved around the chain.

Therefore the Bethe equations are in this case also of the form
{\small\beaa
1   \!\!\!\!&=&\!\!\!\!
\label{baev3I2d7}
 R_0(\xpm_k)^4 \! \left(\tfrac{\xp_k}{\xm_k}\right)^{\!\!2(J-1)}
       \!  \prod_{\ell\neq k}^\KI \!S_0(\xpm_k,\xpm_\ell)^2 S_0(\xpm_k,\mxmp_\ell)^2
\!\prod_{\alpha=1}^2\!\prod_{\ell=1}^{K_{(\alpha)}^\lii}\!\! \sqrt{\tfrac{\xp_k}{\xm_k}}\frac{y_{\ell}^{(\alpha)}\!-\xm_k}{y_{\ell}^{(\alpha)}\!-\xp_k} \sqrt{\tfrac{\xp_k}{\xm_k}}\frac{y_{\ell}^{(\alpha)}\!+\xm_k}{y_{\ell}^{(\alpha)}\!+\xp_k}\\
\label{baev3II2d7}
      1 \!\!\! \!&=&\! \! \!\!\!\! \left(\frac{y_k^{(\alpha)}\!- x_B}{y_k^{(\alpha)}\!+x_B}\right)^2\!
         \prod_{\ell=1}^\KI  \sqrt{\tfrac{\xp_\ell}{\xm_\ell}} \frac{y_{k}^{(\alpha)}\!-\xp_\ell}{y_{k}^{(\alpha)}\!-\xm_\ell} \sqrt{\tfrac{\xm_\ell}{\xp_\ell}}    \frac{y_{k}^{(\alpha)}\!+\xm_\ell}{y_{k}^{(\alpha)}\!+\xp_\ell}
         \prod_{\ell=1}^\KIIIA \frac{w_\ell^{(\alpha)}\!-v_k^{(\alpha)}\! -\frac{i}{2g}}{w_\ell^{(\alpha)}\!-v_k^{(\alpha)}\!+\frac{i}{2g}}
                             \;\; \frac{w_\ell^{(\alpha)}\!+v_k^{(\alpha)}\! +\frac{i}{2g}}{w_\ell^{(\alpha)}\!+v_k^{(\alpha)}\!-\frac{i}{2g}} \\
\label{baev3III2d7}
      1 \!\!\!\!&=&\!
         \prod_{\ell=1}^\KIIA \frac{w_k^{(\alpha)}\!-v_\ell^{(\alpha)} \!+\frac{i}{2g}}{w_k^{(\alpha)}\!-v_\ell^{(\alpha)}\!-\frac{i}{2g}}\;\;
                               \frac{w_k^{(\alpha)}\!+v_\ell^{(\alpha)}\! -\frac{i}{2g}}{w_k^{(\alpha)}\!+v_\ell^{(\alpha)}\!+\frac{i}{2g}}
         \prod_{\ell\neq k}^\KIIIA \frac{w_k^{(\alpha)}\! - w_\ell^{(\alpha)} \!- \frac i g}{w_k^{(\alpha)}\! - w_\ell^{(\alpha)} \!+ \frac i g}
                                \;\; \frac{w_k^{(\alpha)}\! + w_\ell^{(\alpha)}\! - \frac i g}{w_k^{(\alpha)}\! + w_\ell^{(\alpha)}\! + \frac i g}.\eeaa}

where now $R_0$ must be the corresponding D7-brane dressing phase\footnote{One should be able to obtain it imposing crossing symmetry as in the giant graviton cases \cite{ChCo,ABR}.}.

\section{Discussion}
In this paper we have used a nested coordinate Bethe ansatz  to obtain the Bethe equations for excitations on free strings on $AdS_5\times S^5$ with certain open-boundary conditions. In particular we considered strings whose large angular momentum is in the same plane as the angular momentum of the maximal D3-brane (giant graviton) to which it is attached. In section \ref{d7} we derived the Bethe equations for the  closely-related case of an open string ending on a D7-brane.
As usual, these equations are asymptotic in the sense that they characterize the spectrum of strings in the limit of very large angular momentum $J$. An obviously interesting step forward would be the formulation of a Boundary Thermodynamic Bethe Ansatz for these cases, which should encode the spectrum of strings with finite angular momentum.

Studying finite size effects for planar AdS/CFT with open boundaries has some appealing features in comparison with the case of periodic boundary conditions. In the first place, at weak coupling, the  analogues of wrapping effects can show up as early as at 1-loop order. Therefore, explicit verifications of finite size corrections to the asymptotic Bethe ansatz spectrum can be made without daunting higher-loop computations. Secondly, for short operators one can swap the roles between background fields and impurities, thus obtaining inequivalent Bethe ansatz descriptions. Interestingly, their failures due to finite size corrections occur at different orders, and this interplay between alternative points of view allows one to test certain finite size corrections without an explicit perturbative gauge field theory computation. In other words, some aspects of finite size effects should be easier to derive in cases with open boundaries and, for instance, explicit checks beyond the leading L\"uscher approximation should be possible.
~

\textit{Acknowledgments.-- } D.H.C. is funded by the Seventh Framework Programme under grant agreement number PIEF-GA-2008-220702. C.A.S.Y. is supported by a fellowship from the Japan Society for the Promotion of Science.

\subsection*{Appendix: 1-loop Bethe equations}

The 1-loop Hamiltonian in the $su(2)$ sector (for states with chiral fields $Y$ and $Z$ only) is:
\ba
H^{su(2)} \! &=& \!  2g^2 \sum_{l=1}^{L-1}(1 - P_{l,l+1}) + 2g^2 q_1^Z + 2g^2 q_{L}^Z,
\label{h1}
\ea
where $q_l^Z$ is 1 or 0 whether the $l^{\rm th}$ site is occupied by a $Z$ or not. These $L$ sites
do not include the boundary sites. So, there are a $0^{\rm th}$ and a $(L+1)^{\rm st}$ site, both occupied by a $Y$.

Let us consider a single bulk $Y$ impurity in a background of $Z$ fields
\be
|\psi\rangle = \sum_{n=1}^L (e^{ipn} + R^{su(2)}_{\rm L}(p)e^{-ipn}) |n\rangle, \qquad |n\rangle \equiv |Y_{\rm L};Z^{n-1}YZ^{L-n};Y_{\rm R}\rangle.
\ee
This  superposition of left-moving and right-moving spinons is an eigenstate with eigenvalue \\
$8g^2\sin^2(\tfrac{p}{2})+4g^2$, provided the following two conditions are met
\be
R^{su(2)}_{\rm L}(p) = - \frac{1-2 e^{ip}}{1-2 e^{-ip}}, \qquad  e^{2ip(L+1)} = (R^{su(2)}_{\rm L}(p))^2.
\ee

Let us consider now states in the $sl(2)$ sector
\be
|a_0,a_1,\cdots,a_{L+1}\rangle \equiv \epsilon^{i_1,\cdots, i_N}_{j_1,\cdots ,j_N}
Z^{j_1}_{i_1} \cdots Z^{j_{N-1}}_{i_{N-1}}
({\cal D}^{a_1}Z {\cal D}^{a_2}Z\cdots {\cal D}^{a_L}Z)^{j_N}_{i_N}.
\ee
In the boundary sites, $a_0$ and $a_{L+1}$ must be different from zero. The 1-loop Hamiltonian
is that of \cite{CoSi} setting what is defined there as $\alpha$ to zero (for maximal giant graviton boundaries).
The Hamiltonian is conveniently split into bulk and boundary terms
\be
H^{sl(2)} = 2g^2 \mathcal{H}_0 + 2g^2 \sum_{l=0}^{L} \mathcal{H}_{l,l+1}+2g^2 \mathcal{H}_{L+1},
\ee
with
\ba
\mathcal{H}_{l,l+1} |a_l,a_{l+1}\rangle \!\! &=& \!\! \sum_{k=1}^{a_l} \frac{1}{k}(|a_l,a_{l+1}\rangle-|a_l-k,a_{l+1}+k\rangle) +
\sum_{k=1}^{a_{l+1}} \frac{1}{k}(|a_l,a_{l+1}\rangle-|a_l+k,a_{l+1}-k\rangle) \nn
\\
\mathcal{H}_{0} |a_0,\dots\rangle \!\! &=& \!\! \sum_{k=1}^{a_0-1} \frac{1}{k}|a_0,\dots\rangle,
\hspace{2.4cm}
\mathcal{H}_{L+1} |\dots,a_{L+1}\rangle = \sum_{k=1}^{a_{L+1}-1} \frac{1}{k} |\dots,a_{L+1}\rangle.
\nn
\ea
Note that whenever $|0,a_1,\cdots,\rangle$ or $|\cdots,a_{L},0\rangle$ is retrieved, those states
must be taken as identically zero. We now consider a single ${\cal D}$ bulk impurity
\be
|\psi\rangle = \sum_{n=0}^{L+1} (e^{ipn} + R^{sl(2)}_{\rm L}(p)e^{-ipn}) |n\rangle,
\ee
with $n$ in $|n\rangle$  indicating the position of the bulk impurity. For example
$|0\rangle = |2,0,\cdots,0,1\rangle$, $|1\rangle = |1,1,\cdots,0,1\rangle,\ \cdots$,
$|L+1\rangle = |1,0,\cdots,0,2\rangle$. For $|\psi\rangle$ to be eigenstate with eigenvalue $8g^2\sin^2(\tfrac{p}{2})+4g^2$, the following two
conditions are required
\be
R^{sl(2)}_{\rm L}(p) = - \frac{1-2 e^{-ip}}{1-2 e^{ip}}, \qquad  e^{2ip(L+1)} = (R^{sl(2)}_{\rm L}(p))^2.
\ee

Both $R^{su(2)}_{\rm L}$ and $R^{sl(2)}_{\rm L}$ are consistent with the weak coupling limits
of the all-loop reflection factors obtained by Hofman and Maldacena \cite{HMopen}
\ba
R^{su(2)}_{\rm L} \!\!&=&\!\! R_{0{\rm L}}^2 A_{{\rm L}}^2 \sim - \frac{1-2 e^{ip}}{1-2 e^{-ip}} +{\cal O}(g^2),
\\
R^{sl(2)}_{\rm L} \!\! &=&\  R_{0{\rm L}}^2 \ \  \sim - \frac{1-2 e^{-ip}}{1-2 e^{ip}} +{\cal O}(g^2).
\ea
Defining, $K^0=L+2$ the total number of sites in the open chain, the Bethe equations read
\ba
\label{sueq}
\text{for a single bulk particle in the $su(2)$ sector:}
\hspace{1.45cm} \left(\tfrac{\xp}{\xm}\right)^{2(K^0-1)} \!\!&=&\!\! R_{0{\rm L}}^4 A_{{\rm L}}^4,
\\
\text{for a single bulk particle in the $sl(2)$ sector:} \hspace{1.45cm} \left(\tfrac{\xp}{\xm}\right)^{2(K^0-1)} \!\!&=&\!\! R_{0{\rm L}}^4.
\label{sleq}
\ea


\begin{thebibliography}{99}


\bibitem{MZ} J.~A.~Minahan and K.~Zarembo,
  JHEP {\bf 0303} (2003) 013
  [arXiv:hep-th/0212208].

\bibitem{BKS}
N.~Beisert, C.~Kristjansen and M.~Staudacher,
  Nucl.\ Phys.\ B {\bf 664} (2003) 131
  [arXiv:hep-th/0303060].

\bibitem{BS}
 N.~Beisert and M.~Staudacher,
  Nucl.\ Phys.\ B {\bf 670} (2003) 439
  [arXiv:hep-th/0307042].

\bibitem{AFS}
  G.~Arutyunov, S.~Frolov and M.~Staudacher,
  JHEP {\bf 0410} (2004) 016
  [arXiv:hep-th/0406256].

\bibitem{BKSZ}
  N.~Beisert, V.~A.~Kazakov, K.~Sakai and K.~Zarembo,
  Commun.\ Math.\ Phys.\  {\bf 263} (2006) 659
  [arXiv:hep-th/0502226].

\bibitem{BSpsu224}
  N.~Beisert and M.~Staudacher,
  Nucl.\ Phys.\  B {\bf 727} (2005) 1
  [arXiv:hep-th/0504190].


\bibitem{Bsu22}
  N.~Beisert
   Adv.\ Theor.\ Math.\ Phys.\  {\bf 12} (2008) 945 [arXiv:hep-th/0511082].


\bibitem{boundstates}
  M.~de Leeuw,
  JHEP {\bf 0806} (2008) 085
  [arXiv:0804.1047 [hep-th]].

  G.~Arutyunov and S.~Frolov,
  Nucl.\ Phys.\  B {\bf 804} (2008) 90
  [arXiv:0803.4323 [hep-th]].


  A.~Torrielli,
  J.\ Phys.\ A  {\bf 42} (2009) 055204
  [arXiv:0806.1299 [hep-th]].

  G.~Arutyunov, M.~de Leeuw and A.~Torrielli,
  JHEP {\bf 0905} (2009) 086
  [arXiv:0903.1833 [hep-th]].

T.~Matsumoto and S.~Moriyama,
  JHEP {\bf 0909} (2009) 097
  [arXiv:0902.3299 [hep-th]].



\bibitem{Janik}
  R.~A.~Janik,
  Phys.\ Rev.\ D {\bf 73} (2006) 086006
  [arXiv:hep-th/0603038].


\bibitem{BHL}
  N.~Beisert, R.~Hernandez and E.~Lopez,
  JHEP {\bf 0611} (2006) 070
  [arXiv:hep-th/0609044].


\bibitem{BES}
  N.~Beisert, B.~Eden and M.~Staudacher,
  J.\ Stat.\ Mech.\  {\bf 0701} (2007) P021
  [arXiv:hep-th/0610251].


\bibitem{TBA}
  G.~Arutyunov and S.~Frolov,
  JHEP {\bf 0903} (2009) 152
  [arXiv:0901.1417 [hep-th]].

  N.~Gromov, V.~Kazakov and P.~Vieira,
  arXiv:0901.3753 [hep-th].

  D.~Bombardelli, D.~Fioravanti and R.~Tateo,
  J.\ Phys.\ A  {\bf 42} (2009) 375401
  [arXiv:0902.3930 [hep-th]].

  N.~Gromov, V.~Kazakov, A.~Kozak and P.~Vieira,
  arXiv:0902.4458 [hep-th].

  G.~Arutyunov and S.~Frolov,
  JHEP {\bf 0905} (2009) 068
  [arXiv:0903.0141 [hep-th]].


\bibitem{HMopen}
  D.~M.~Hofman and J.~M.~Maldacena,
 JHEP {\bf 0711} (2007) 063
  [arXiv:0708.2272 [hep-th]].


\bibitem{BV}
  D.~Berenstein and S.~E.~Vazquez,
  JHEP {\bf 0506} (2005) 059
  [arXiv:hep-th/0501078].

\bibitem{Mann}
  N.~Mann and S.~E.~Vazquez,
  JHEP {\bf 0704} (2007) 065
  [arXiv:hep-th/0612038].




\bibitem{ChCo}
  H.~Y.~Chen and D.~H.~Correa,
  JHEP {\bf 0802} (2008) 028
  [arXiv:0712.1361 [hep-th]].

\bibitem{ABR}
  C.~Ahn, D.~Bak and S.~J.~Rey,
  JHEP {\bf 0804} (2008) 050
  [arXiv:0712.4144 [hep-th]].


\bibitem{Ahn:2008df}
  C.~Ahn and R.~I.~Nepomechie,
  JHEP {\bf 0805} (2008) 059
  [arXiv:0804.4036 [hep-th]].

\bibitem{Murgan:2008zu}
  R.~Murgan and R.~I.~Nepomechie,
  JHEP {\bf 0806} (2008) 096
  [arXiv:0805.3142 [hep-th]].

\bibitem{Beisert:2008cf}
  N.~Beisert and F.~Loebbert,
  Adv.\ Sci.\ Lett.\  {\bf 2}, 261 (2009)
  [arXiv:0805.3260 [hep-th]].

\bibitem{Palla}
  L.~Palla,
  Nucl.\ Phys.\  B {\bf 808}, 205 (2009)
  [arXiv:0807.3646 [hep-th]].



\bibitem{Galleas}
  W.~Galleas,
  Nucl.\ Phys.\  B {\bf 820}, 664 (2009)
  [arXiv:0902.1681 [hep-th]].

\bibitem{Nepomechie:2009zi}
  R.~I.~Nepomechie,
  JHEP {\bf 0905} (2009) 100
  [arXiv:0903.1646 [hep-th]].

\bibitem{KMMW}
  M.~Kruczenski, D.~Mateos, R.~C.~Myers and D.~J.~Winters,
  JHEP {\bf 0307} (2003) 049
  [arXiv:hep-th/0304032].



\bibitem{ErMa}
  T.~Erler and N.~Mann,
  JHEP {\bf 0601}, 131 (2006)
  [arXiv:hep-th/0508064].

\bibitem{CY}
  D.~H.~Correa and C.~A.~S.~Young,
  J.\ Phys.\ A  {\bf 41} (2008) 455401
  [arXiv:0808.0452 [hep-th]].


\bibitem{Beis2006}
  N.~Beisert,
  J.\ Stat.\ Mech.\  {\bf 0701} (2007) P017
  [arXiv:nlin/0610017].


\bibitem{LLM}
  H.~Lin, O.~Lunin and J.~M.~Maldacena,
  JHEP {\bf 0410} (2004) 025
  [arXiv:hep-th/0409174].
\bibitem{HM}
 D.~M.~Hofman and J.~M.~Maldacena,
 J.\ Phys.\ A {\bf 39}, 13095 (2006) [arXiv:hep-th/0604135].

\bibitem{gaudin} M. Gaudin,
Phys. Rev. {\bf A4} (1971) 386.


\bibitem{olpe} M.A. Olshanetsky and A.M. Perelomov,
Phys. Reports \textbf{94} (1983) 313.

\bibitem{che}
I.~V.~Cherednik,
Theor. Math. Phys. \textbf{61} (1984) 977.



\bibitem{Sutherland}
B.~Sutherland,
J.\ Math.\ Phys.\  {\bf 21} (1980) 1770.


\bibitem{Sklyanin}
  E.~K.~Sklyanin,
  J. Phys. A: Math. Gen. {\bf 21} (1988) 2375


\bibitem{FaddeevABA}
  L.~D.~Faddeev,
  arXiv:hep-th/9605187.


\bibitem{schulz}
H.~Schulz,
J. Phys. C: Solid State Phys. {\bf 18} (1985) 581


\bibitem{yang}
  C.~N.~Yang,
  Phys. Rev. Lett {\bf 19} (1967) 1312.


\bibitem{AFZ}
  G.~Arutyunov, S.~Frolov and M.~Zamaklar,
  JHEP {\bf 0704} (2007) 002
  [arXiv:hep-th/0612229].

\bibitem{AF2}
  G.~Arutyunov and S.~Frolov,
  JHEP {\bf 0712}, 024 (2007)
  [arXiv:0710.1568 [hep-th]].

\bibitem{deLeeuw}
  M.~de Leeuw,
  J.\ Phys.\ A  {\bf 40} (2007) 14413
  [arXiv:0705.2369 [hep-th]].


\bibitem{Martins}
  M.~J.~Martins and C.~S.~Melo,
  Nucl.\ Phys.\  B {\bf 785} (2007) 246
  [arXiv:hep-th/0703086].


\bibitem{B2003}
 N.~Beisert,
 Nucl.\ Phys.\  B {\bf 682} (2004) 487
 [arXiv:hep-th/0310252].


\bibitem{deLeeuw:2008}
  M.~de Leeuw,
  JHEP {\bf 0901} (2009) 005
  [arXiv:0809.0783 [hep-th]].

\bibitem{CY2}
  D.~H.~Correa and C.~A.~S.~Young,
  JHEP {\bf 0908} (2009) 097
  [arXiv:0905.1700 [hep-th]].

\bibitem{CoSi}
  D.~H.~Correa and G.~A.~Silva,
   JHEP {\bf 0611} (2006) 059
  [hep-th/0608128].



\end{thebibliography}
\end{document}